\documentclass[sigconf]{acmart}
\usepackage{xspace}
\usepackage{graphicx}
\usepackage{xcolor}
\usepackage{xargs}
\usepackage{enumitem}
\usepackage{verbatim}

\usepackage{amsmath}  
\usepackage{placeins}  
\usepackage{array}

\usepackage{multirow}
\usepackage{subcaption}
\usepackage{arydshln}
\usepackage{twemojis}
\usepackage[table]{xcolor}

\usepackage{listings}

\usepackage{pgfplots}
\pgfplotsset{compat=1.18}

\usepackage[utf8]{inputenc}
\newcommand{\td}[1] 
{{\color{red} $\leftarrow$ {\bf TO DO}: #1 }}

\usepackage{xcolor}
\usepackage{soulutf8}         
\definecolor{quotegrey}{RGB}{242,242,242}
\sethlcolor{quotegrey}

\newcommand{\eg}{{\it e.g.,\ }}
\newcommand{\etal}{{\it et al.\ }}

\newcommand{\ie}{{\it i.e.,\ }}

\newcommand{\tool}{\textit{Personagram}\xspace}

\newlength{\tboxsize}
\lstdefinestyle{prompt}{
  basicstyle=\ttfamily\small,
  breaklines=true,
  frame=single,
  backgroundcolor=\color{gray!10},
}

\definecolor{red}{RGB}{178,34,34}

\newif\iftrack
\tracktrue 
\newcommandx{\revised}[1][]{\iftrack{\textcolor{black}{#1}}\else#1\fi}

\AtBeginDocument{%
  }

\begin{document}

\title{Personagram: Bridging Personas and Product Design for Creative Ideation with Multimodal LLMs}






\author{Taewook Kim}
\affiliation{%
  \institution{Toyota Research Institute}
  \city{Los Altos}
  \state{CA}
  \country{USA}}
\email{taewook@u.northwestern.edu}

\author{Matthew K. Hong}
\affiliation{%
  \institution{Toyota Research Institute}
  \city{Los Altos}
  \state{CA}
  \country{USA}}
\email{matt.hong@tri.global}

\author{Yan-Ying Chen}
\affiliation{%
  \institution{Toyota Research Institute}
  \city{Los Altos}
  \state{CA}
  \country{USA}}
\email{yan-ying.chen@tri.global}

\author{Jonathan Q. Li}
\affiliation{%
  \institution{Toyota Research Institute}
  \city{Los Altos}
  \state{CA}
  \country{USA}}
\email{jonathan.li@tri.global}

\author{Monica P Van}
\affiliation{%
  \institution{Toyota Research Institute}
  \city{Los Altos}
  \state{CA}
  \country{USA}}
\email{monica.van@tri.global}

\author{Shabnam Hakimi}
\affiliation{%
  \institution{Toyota Research Institute}
  \city{Los Altos}
  \state{CA}
  \country{USA}}
\email{shabnam.hakimi@tri.global}

\author{Matthew Kay}
\affiliation{%
  \institution{Northwestern University}
  \city{Evanston}
  \state{IL}
  \country{USA}}
\email{matthew.kay@u.northwestern.edu}

\author{Matthew Klenk}
\affiliation{%
  \institution{Toyota Research Institute}
  \city{Los Altos}
  \state{CA}
  \country{USA}}
\email{matt.klenk@tri.global}

\renewcommand{\shortauthors}{Kim et al.}

\begin{abstract}
Product designers often begin their design process with handcrafted personas. While personas are intended to ground design decisions in consumer preferences, they often fall short in practice by remaining abstract, expensive to produce, and difficult to translate into actionable design features. As a result, personas risk serving as static reference points rather than tools that actively shape design outcomes. To address these challenges, we built Personagram, an interactive system powered by multimodal large language models (MLLMs) that helps designers explore detailed census-based personas, extract product features inferred from persona attributes, and recombine them for specific customer segments. In a study with 12 professional designers, we show that Personagram facilitates more actionable ideation workflows by structuring multimodal thinking from persona attributes to product design features, achieving higher engagement with personas, perceived transparency, and satisfaction compared to a chat-based baseline. We discuss implications of integrating AI-generated personas into product design workflows.

%




\end{abstract}

\ccsdesc[500]{Human-centered computing~Interactive systems and tools}
\ccsdesc[500]{Human-centered computing~Human computer interaction (HCI)}


\keywords{Creativity support, Multimodal LLMs, Persona, Product design}

\begin{teaserfigure}
    \includegraphics[width=\textwidth]{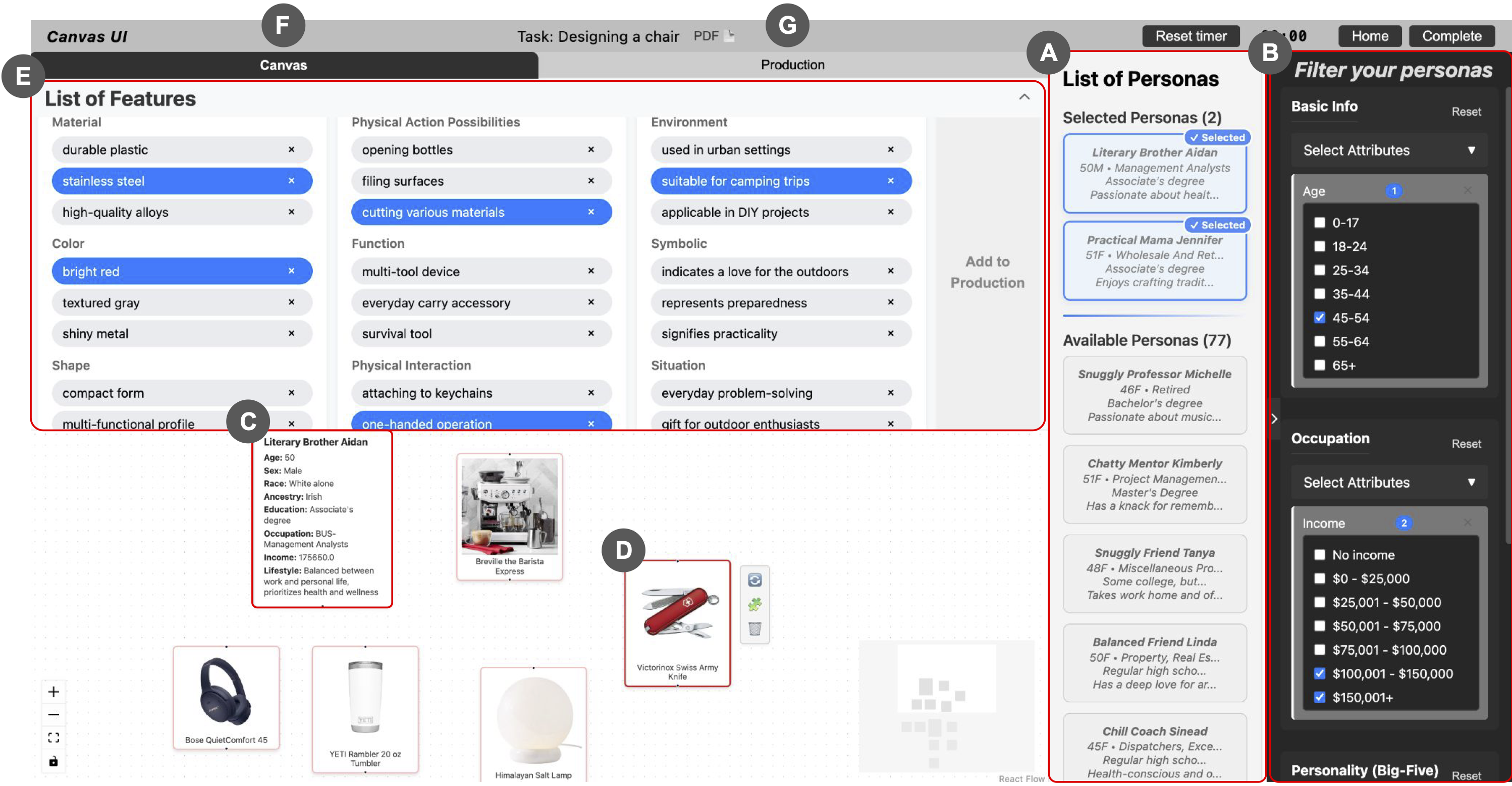}
    \caption{Personagram system: an AI-based tool enabling persona-driven product design. Main system components include A) Persona List, B) Attribute Filter, C) Persona Tile, D) Product Reference Image, E) Design Feature List, F) Canvas UI (Current), and G) Production UI (Prompt Interface)}
    \Description{A screenshot of the Personagram software interface, labeled with components A through G. The central workspace, labeled "Canvas UI" (F), contains "Persona Tiles" (C) displaying demographic profiles and "Product Reference Images" (D), such as a Swiss Army knife. Above the canvas, a "List of Features" (E) panel organizes design attributes like Material, Color, and Function. The right-hand side features a "List of Personas" (A) showing selected and available profiles, adjacent to an "Attribute Filter" (B) sidebar with checkboxes for demographic filtering. Tabs at the top allow switching between the current Canvas view and a "Production" (G) interface.}
\label{fig:teaser}
\end{teaserfigure}


\maketitle

\section{Introduction}
Personas are a longstanding tool in product design practice, intended to anchor design decisions in user needs and preferences~\cite{cooper1999inmates, pruitt2010persona}. By synthesizing demographic and behavioral data into archetypal profiles, personas are meant to help teams imagine the perspectives of prospective users and ground product concepts in real-world contexts. 

Yet in practice, personas often fall short of these goals. They are expensive and time-consuming to produce, heavily reliant on qualitative synthesis, and, due to their static nature, quickly become outdated as markets and consumer preferences evolve. Their abstract narratives hardly translate into actionable product requirements, leaving designers with little guidance on how persona attributes should inform concrete features of the eventual product~\cite{matthews2012designers}. In many organizations, personas risk being reduced to performative exercises, or \textit{persona theater}, where teams showcase polished persona posters or slide decks without integrating them into real design work. As a result, personas frequently become static artifacts referenced at the start of a project, and then sidelined, rather than active tools that structure ongoing design decision-making.

This disconnect has serious consequences. When personas fail to directly inform design features, teams risk producing products that reflect internal biases (e.g., self-referential) more than consumer realities~\cite{li2024chatgpt,emmanuel2022related}. Design rationales become less interpretable, weakening the traceability from personas to product outcomes. In industries with long development cycles, such as automotive and robotics, this gap can lead to costly\footnote{For instance, automotive industries conduct multiple consumer clinics that cost over \$100,000 each time to evaluate and reduce this mismatch~\cite{burnap2023product}.} mismatches between consumer expectations and designed products in ways that undermine the very purpose of human-centered design. Importantly, this breakdown is not merely a failure of persona fidelity or data quality, but of interaction design.

Prior HCI research and recent LLM-powered commercial tools\footnote{\url{https://persona.design/} \\ \url{https://uxpressia.com/ai-persona-generator}} explored methods for improving the use of personas by making them more engaging or lifelike by, for instance, anthropomorphizing them into ``talking'' personas or adding immersive elements that make personas feel more present~\cite{kaate2025personas,sun2024building,amin2025generative,zhou2025vivid,bonnardel2020enhancing}. These approaches predominantly adopt a chat-based interaction paradigm, where designers query personas through free-form dialogue and receive open-ended responses. While conversational flexibility lowers the barrier to engagement, it places the burden of structuring design reasoning on the designer as they translate personifying details to concrete product features. Ultimately, these approaches stop short of solving the more fundamental challenge: \textit{how can we reimagine personas into tools that actively shape design decisions?} 

Significant recent advances in artificial intelligence provide new opportunities to overcome these limitations. Large language models (LLMs) can generate personas that are internally consistent, richly detailed, and grounded in population-level statistics such as census records or opinion polls~\cite{kirk2024prism,wang2025deeppersona,castricato2024persona}. These AI-augmented personas offer designers a means of speculating about user needs and preferences even in contexts where direct user feedback is unavailable, such as in early stage concept development or industries with limited access to prospective users\cite{amin2025generative,chen2024empathy}. Moreover, multimodal LLMs (MLLMs) \cite{hertz2024stylealigned,ye2023ip-adapter} integrate reasoning across text and imagery, enabling designers to translate across visual inspiration and concept articulation. We leverage these capabilities to open new pathways for augmenting cognitive processes for design speculation, helping designers move beyond static, descriptive personas toward dynamic and steerable tools that directly connect consumer attributes with design features.

In this paper, we introduce \tool, an interactive system that operationalizes these advances in AI to more directly connect personas with product design outcomes. \tool enables designers to explore richly detailed, census-based personas, infer product features from persona attributes, and incorporate those features into new products. The system builds on the idea of product references---examples that illustrate how different individuals might perceive or relate to a design~\cite{crilly2004seeing}—and leverages the vast internet-scale corpus of such references available through MLLMs. By linking persona attributes to visualizable design elements, \tool transforms personas from abstract storytelling devices (largely driven by free-form dialogue) into interactive tools for structuring design ideation by decomposing persona-driven ideation into multi-round, multimodal steps. Through a within-subjects study with 12 professional designers, we demonstrate how \tool supports more engaging ideation workflows compared to a Persona + chat-based interaction baseline. Specifically, participants using \tool reported higher levels of satisfaction when exploring ideas (in shorter time), engaged more frequently with persona-driven outputs (product references vs. free-form LLM responses), and spent less time refining prompts. They also perceived the \tool system as being more transparent, attributing this to the structured text$\rightarrow$image$\rightarrow$text$\rightarrow$image multimodal system pipeline that scaffolded the ideation process from viewing personas to prototyping solutions. Finally, participants reported on the perceived value of tool for professional practice, noting its potential for saving time and providing support in establishing alignment with design stakeholders. 

We contribute the following:
\begin{itemize}
    \item An MLLM pipeline that connects persona attributes to product features by inferring relevant products, retrieving image representations, and extracting textual design features. 
    \item An interactive system, \tool, that operationalizes this pipeline to support persona-driven structured ideation workflows by enabling designers to dynamically explore persona-relevant product references and scaffold the extraction, recombination, and refinement of product features into novel concept directions.
    \item A within-subjects evaluation with 12 professional designers, showing that \tool enables more actionable ideation workflows than a Persona + chat-based baseline, with participants reporting higher satisfaction, more frequent engagement with persona-driven outputs, reduced prompt refinement effort, and greater perceived system transparency. 
\end{itemize}

\section{Related Work}
\label{rw}
\subsection{Models of Consumer Product Experience}
Developing tools to amplify product designers requires deep insight into the underlying human cognitive processes through which consumers respond to product design~\cite{reber2004processing}. Most prominently, Crilly \etal provide a framework of consumer response to the visual domain in product design, which identifies three cognitive processes (informed by prior work such as~\cite{crozier1994manufactured,cupchik1999emotion, lewalski1988product,norman2007emotional}) through which consumers interpret product form~\cite{crilly2004seeing}. \textit{Aesthetic} impression refers to the immediate sensory response to a product's visual features (\eg shape, color, and texture), \textit{semantic} interpretation involves inferring functional or behavioral information from the product's form (\eg a handle suggests gripping), and \textit{symbolic} association captures the personal or social significance attached to design, allowing users to associate it with identity, lifestyle, or values (\eg Rolex watch signals wealth). 

Their framework also emphasizes the importance of understanding the visual references consumers draw upon---features that often resemble stored visual memories---when encountering new products~\cite{asuzu2023towards}. For example, a consumer familiar with precisely machined aluminum unibody Apple devices may associate similar aesthetic cues in an unfamiliar product (\eg a smart home thermostat) with premium quality and thoughtful design. Such associations can enhance the perceived desirability of new products even in the absence of brand recognition. Supporting this perspective, Reber \etal and Martindale argued that visual or semantic familiarity can increase judgments of aesthetic pleasure, reinforcing the role of product references in shaping perception~\cite{reber2004processing, martindale1984pleasures}. Recent computational approaches have begun to operationalize this perspective by modeling how consumers rely on such references when evaluating novel products. For instance, Kobayashi and Takeda validated a product recommendation model using consumers' e-commerce data, showing that preferences for new products closely matched aesthetic details, such as color and material~\cite{kobayashi2021product}, of their previously favorited products.

Because designers often curate reference images in their practice to inspire new concepts~\cite{jagtap2017inspiration,gonccalves2014inspires}, Crilly \etal speculated on the potential benefits of systematically generating these references as a way to bridge the gap between designer intent and consumer perception. The authors identify six categories that highlight the diverse ways references can shape meaning: stereotypes, similar products, metaphors (cross-domain analogies), characters, conventions, and clichés. Extending this idea, Asuzu and Olechowski further argued that these visual product references (hereafter, \textit{product references}) not only shape consumer perception but also present an opportunity for design tools to proactively guide and influence aesthetic judgment during early-stage concept development~\cite{asuzu2023towards}. 

While Crilly's model is primarily used as an analytical lens, our work builds on its structure by leveraging LLMs' broad knowledge of consumer products to generate the kinds of product references or features that might appear familiar to the personas as well as their cognitive responses to them. Our approach reinterprets the model as a generative framework that enables the creation of ideation tools capable of generating product concepts through persona-specific aesthetic, semantic, and symbolic interpretations. To realize this interpretation of Crilly's model, we draw on two complementary advances to connect prospective personas to concrete product concepts and close the gap: 1) multimodal large language models (MLLMs) that enable the identification, decomposition, and recombination of product features across modalities, and 2) interactive AI personas that capture possible reference products to support early-stage ideation.


\subsection{Creative Decomposition and Recombination with Multimodal LLMs}
Multimodal large language models (MLLMs) have increasingly been leveraged to decompose and recombine design features from reference images to enhance visual exploration and generation~\cite{vinker2023concept}. Building on these latest advances, recent systems in HCI research embed MLLMs within interactive tools that guide designers in exploring visual concept and style as well as refining multimodal prompts and visual outputs. For instance, the S\&UI system extracts semantic content from interface screenshots to support exploratory search of UI interface design ideas~\cite{park2025leveraging}. Beyond exploration, CreativeConnect leveraged semantic content extraction from reference images such as subject matter, action, and theme to scaffold later recombination of extracted keywords to use in a text-to-image (T2I) generation pipeline~\cite{choi2024creativeconnect}. AutoSpark combined semantic prompt scaffolding with kansei engineering to search corresponding car images with abstract keywords like elegant, which are then translated into shape, color and texture features to be used in T2I generation~\cite{chen2024autospark}. Other work such as FusAIn~\cite{peng2025fusain} and Brickify~\cite{shi2025brickify} blend subject and style feature extraction with direct manipulation to specify and refine design intent through multimodal prompting~\cite{shi2025brickify}.

To enable greater facilitation of exploration and generation, other systems employed conceptual metaphors to transfer aesthetic and functional features into the target design context. ProductMeta~\cite{zhou2025productmeta} advances metaphor-driven ideation by first decomposing the design problem according to Norman's three levels of emotional experience---visceral, behavioral, and reflective~\cite{norman2007emotional}---to guide inspiration discovery. The system then maps features from metaphorical sources to the problem domain, and structures resulting design concepts using Gero’s Function-Behavior-Structure (FBS) model~\cite{gero1990design} alongside aesthetic features such as color, material, and finish. Similarly, DesignFusion~\cite{chen2024designfusion} supports problem decomposition using the 5W1H (who, what, where, when, why, how) framework, and facilitates conceptual exploration and synthesis by exposing the AI model's internal reasoning through both FBS-based representations and extracted aesthetic features. Finally, DesignFromX~\cite{duan2025designfromx} allows users to analyze, select, and integrate design features from across searched reference product images to scaffold intuitive exploration of the design space to shape outcomes that align with individual preferences. 

In summary, the active area of research integrating MLLMs into design workflows highlights our gaps in understanding how different approaches impact the design process. In our work, we show that further support could be achieved through the use of MLLMs that tie persona attributes to corresponding sets of product references, which then serve as grounding material for ideation. By linking user-centered ideation processes with visual exemplars, our approach provides designers with a structured yet adaptable means to navigate the design space, and generate novel product concepts that remain aligned with intended user profiles.

\subsection{Evolving Role of Personas in Design Practice}
A common starting point for product design practice involves creating personas---"fictitious, specific, concrete representations of target users"---that help designers internalize user needs, preferences, and values~\cite{cooper1999inmates,pruitt2010persona}. While personas have long been promoted as a means to help designers focus on user goals and align team decisions throughout the user-centered design process~\cite{miaskiewicz2011personas}, empirical studies suggest a disconnect between the theoretical promise of personas and their actual uptake in real-world design settings---personas are often underutilized and ineffectively applied in practice~\cite{matthews2012designers}. Matthews \etal found that UX practitioners primarily used personas as a communication aid instead of a design aid, viewing personas as too abstract, impersonal, misleading and even distracting to inform design decisions~\cite{matthews2012designers}. Salminen \etal report that persona usage in industry is scattered and lacking clear guidelines beyond general claims of benefits~\cite{salminen2022use}. While the authors also report that personas are often referenced at various stages of design---such as conceptualization, brainstorming, and prototyping---their application typically serves more as a superficial reminder of the consumer voice than as a tool that substantively shapes design decisions.

Emerging research on AI-generated personas and agent-based simulations highlights promising opportunities to support product design practice by approximating human-like interactions, social dynamics, and usability feedback. Kaate \etal show that LLM-based personas offer real-time access to comprehensive, interactive user models that people find engaging and coherent, signaling opportunities for developing deep insight into user behavior and needs and enhanced ideation~\cite{kaate2025you, kaate2025personas}. Tools such as Social Simulacra~\cite{park2022social} and AgentSociety~\cite{piao2025agentsociety} offer LLM-powered prototyping techniques that generate large-scale, realistic social interactions to help designers anticipate social behaviors in interactive systems. Vivid-Persona is a customizable persona tool that integrates generative models with real-time interactive video chat and emotional expressiveness, addressing limitations of static personas such as insufficient interactivity, limited empathy, and coarse user segmentation\cite{zhou2025vivid}. Other research systems employ LLM-generated audience and expert personas to guide creative ideation in writing~\cite{benharrak2024writer}, research~\cite{liu2024personaflow}, and video creation~\cite{choi2025proxona} by simulating perspectives that mirror real-world stakeholders. Expanding this line of work into evaluation, recent research simulate diverse users interacting with mobile or web applications such as ecommerce websites to automatically generate usability feedback, enabling early-stage iteration without human participants~\cite{xiang2024simuser,bougie2025simuser,lu2025uxagent,wang2025opera}. Complementing these applications, Canvil examines how designers adapt LLM outputs into user experiences, foregrounding the role of designerly interpretation in AI-human collaboration~\cite{feng2025canvil}.

These studies demonstrate growing momentum toward using AI personas not only to represent users, but also to simulate feedback, spark ideation, and prototype social interaction—expanding the functional role of personas in product design practice. While prior work highlights the potential utility of personas in specific design activities, we still lack an understanding of how they can support designers across multiple stages of ideation. To address this knowledge gap, we developed a MLLM-powered interactive system that embed personas directly into concept ideation and visual design workflows, thereby allowing designers to navigate the space of possible personas and to identify key reference products rather than treating personas simply as static artifacts.

\section{Design Goals}
Our overarching aim is to enable product designers to use personas not merely as static artifacts, but as active tools that directly shape design outcomes. To achieve this, we designed \tool with the goal of scaffolding user-centered design processes that typically demand significant cognitive effort, which include synthesizing user research insights into personas, guiding concept generation, and informing the prototyping of design solutions~\cite{salminen2022use}.
These considerations informed the following system design goals:  

\begin{itemize}  
    \item \textbf{Design Goal 1. Provide efficient means to explore the persona space (DG1)}
    Personas are commonly used to ground design in user-centered perspectives, yet designers lack efficient mechanisms to explore diverse persona spaces beyond a small set of manually crafted profiles. To address this, we began with a representative dataset of 1,000 personas from which designers can discover, preview, drill down, and select as starting points for ideation. This supports breadth of exploration while reducing the overhead of manually creating personas for early-stage design.  
    
    \item \textbf{Design Goal 2. Surface design dimensions and features from persona-driven product space. (DG2)}  
    A critical bottleneck in design practice lies in moving from high-level persona attributes to actionable design features. Drawing on Crilly \etal’s framework of consumer response to product design~\cite{crilly2004seeing}, we leveraged MLLMs to infer product preferences from persona demographics and behaviors, thereby automatically generating reference products. We further appropriated the \textit{aesthetic}, \textit{semantic}, and \textit{symbolic} cognitive pathways to organize salient design features from these products, surfacing design dimensions tied directly to persona needs and preferences.  

    \item \textbf{Design Goal 3. Support rapid iterative prototyping with identified product features. (DG3)}  
    Designers frequently iterate by experimenting with different combinations of product features, yet this process is time-consuming and cognitively burdensome without tooling support. To facilitate this, we extend extracted features into a dimension scaffolding approach (building on ~\cite{tao2025designweaver}) that transforms independent features into natural language descriptions of novel concepts. These descriptions can then serve as inputs to text-to-image generation models, enabling rapid creation and iteration of visual prototypes that embody persona-driven design directions.  

    \item \textbf{Design Goal 4. Streamline transitions across ideation stages. (DG4)}  
    Designers often expend considerable cognitive effort when transitioning across ideation stages—from persona exploration to feature identification to prototyping. To reduce this friction, we introduced structured scaffolds that chain outputs from one stage (e.g., persona attributes, reference products, feature sets) directly into the next, enabling fluid transitions, and supporting continuous, iterative concept development.
\end{itemize}

\section{Personagram: Bridging Personas and Product Design Ideation by Multimodal LLM}

\subsection{Persona Data}
Personas play a crucial role in our system by inspiring diverse product design ideas. To ensure a wide range of perspectives, we incorporated the PERSONA dataset\footnote{\url{https://huggingface.co/datasets/SynthLabsAI/PERSONA}} into our system. This dataset comprises 1,000 procedurally generated personas that reflect diverse user profiles based on U.S. Census data, thereby supporting improved pluralistic alignment~\cite{castricato2024persona}. Each persona consists of detailed objective and subjective attributes. As defined in previous work~\cite{li2025llm}, objective attributes cover objectively verifiable characteristics such as age, income, sex, marital status, and education level, while subjective ones include lifestyle, personality traits, and ideological beliefs (see Appendix~\ref{appx:attribute}). Because the raw dataset contains 38 distinct attributes, we created a higher-level category breakdown based on formative discussions with three professional product designers (one of whom is a co-author). We merged redundant categories and removed attributes less relevant to early-stage design, including, but not limited to, health insurance, political ideology, religion, citizenship, spoken language, and race.

\revised{We acknowledge that accurately representing human personas via AI is an open challenge and that current representations are not fully solved problems~\cite{santurkar2023whose}. However, the primary goal of this study is not to validate the perfect fidelity of AI-simulated personas, but to explore how recent advancements in pluralistic alignment can be utilized to scaffold the product design process. In professional design practice, personas serve less as statistically precise simulations and more as ``empathy proxies'' or tools to ground ideation~\cite{pruitt2003personas, matthews2012designers}. Therefore, we posit that absolute representational precision is not a strict prerequisite for utility in this domain; rather, the system's ability to offer structured, diverse, and distinct viewpoints is the main goal to support designers in exploring the idea space.}

\begin{figure*}
  \includegraphics[width=\textwidth]{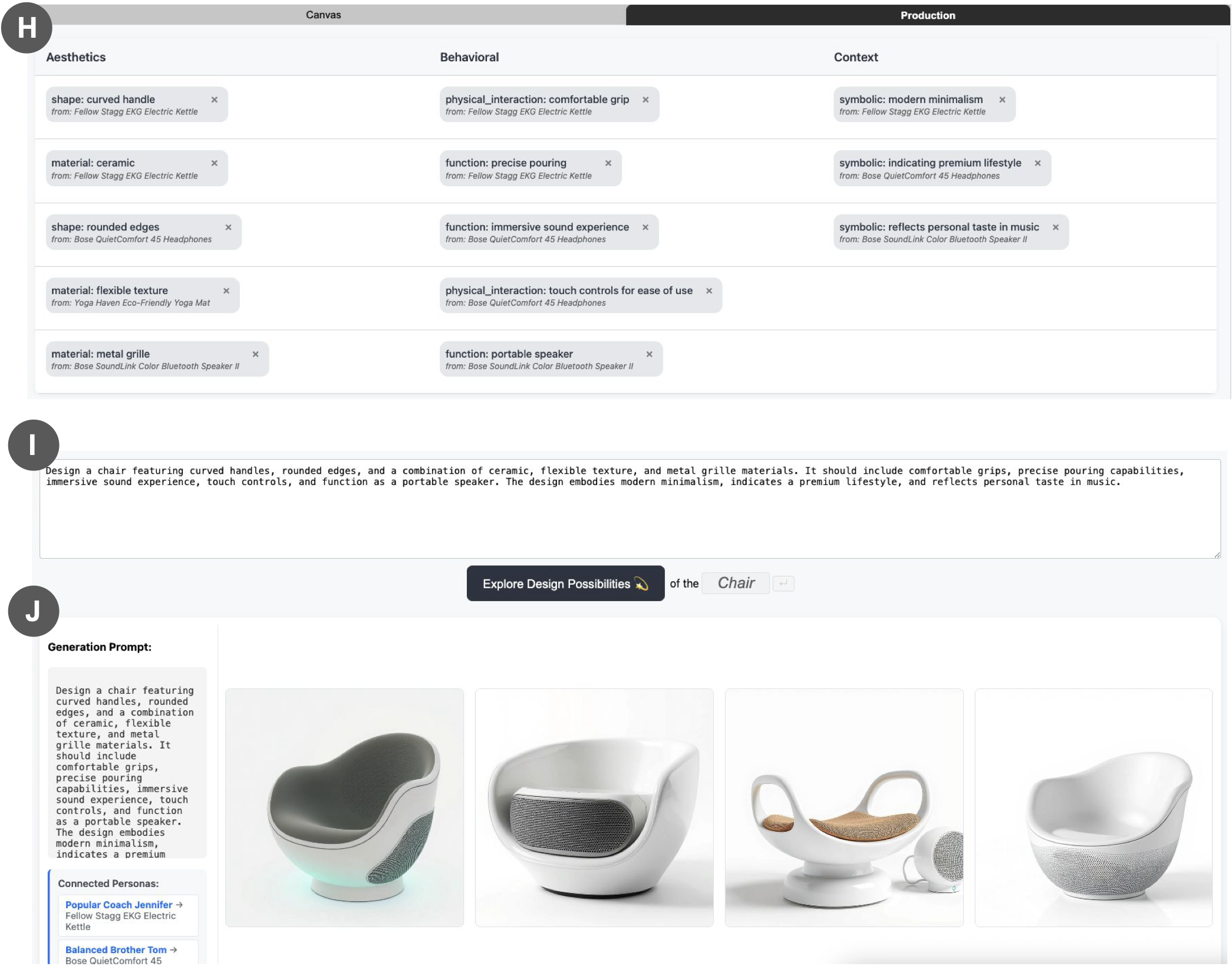}
  \caption{Personagram Production UI Components. Features identified from the Canvas UI can be modified in the H) Dimensional Scaffolding Interface, with changes to the feature list being reflected on the I) Text-to-image Prompt Input Box in real time. Pressing the "Explore Design Possibilities" button triggers image generation and results are displayed on the J) Image Gallery.}
  \label{fig:production}
  \Description{A screenshot of the Personagram Production UI, divided into three vertical sections labeled H, I, and J. The top section (H), the "Dimensional Scaffolding Interface," displays design feature cards (e.g., "shape: curved handle," "material: ceramic") organized into columns for Aesthetics, Behavioral, and Context. The middle section (I) contains a "Text-to-image Prompt Input Box" populated with a natural language description synthesized from the features above, alongside an "Explore Design Possibilities" button. The bottom section (J), the "Image Gallery," displays four AI-generated images of a modern, white chair incorporating speaker-like mesh elements, with a sidebar on the left showing the generation prompt and connected personas.}
\end{figure*}

\subsection{Interface Design and Features}

Figure \ref{fig:teaser} illustrates our interface design, including three important features: persona exploration, canvas UI and production UI.

\subsubsection{Persona Exploration}
The persona exploration area spans two vertical column panels on the right most side, including Persona List (Fig. \ref{fig:teaser}A), and Attribute Filter (Fig. \ref{fig:teaser}B). In this space, users can browse through and explore a list of 1000 personas, preview a detailed breakdown (upon mouse hover), and filter personas based on demographic and behavioral attribute categories (\textbf{DG1}). Persona attributes are hierarchically organized, starting with Basic Info, Occupation, Personality, and Behavioral Traits, From each category, users can access a dropdown list which displays a more granular attribute filter. Exact breakdown of this hierarchy is described in Appendix (\ref{appx:attribute}).

\subsubsection{Canvas UI: Concept Exploration}
The Canvas UI view (Fig. \ref{fig:teaser}F) is a blank canvas akin to digital moodboarding tools (e.g., Miro Board) where users can import images, attach post it notes, pan/zoom-in/zoom-out and interact with them through drag and drop interactions to facilitate creative ideation. Upon selecting a persona from the list of personas, a Persona Tile (Fig. \ref{fig:teaser}C) is created and displayed on the canvas. 

\begin{figure*}[t!]
  \includegraphics[width=\textwidth]{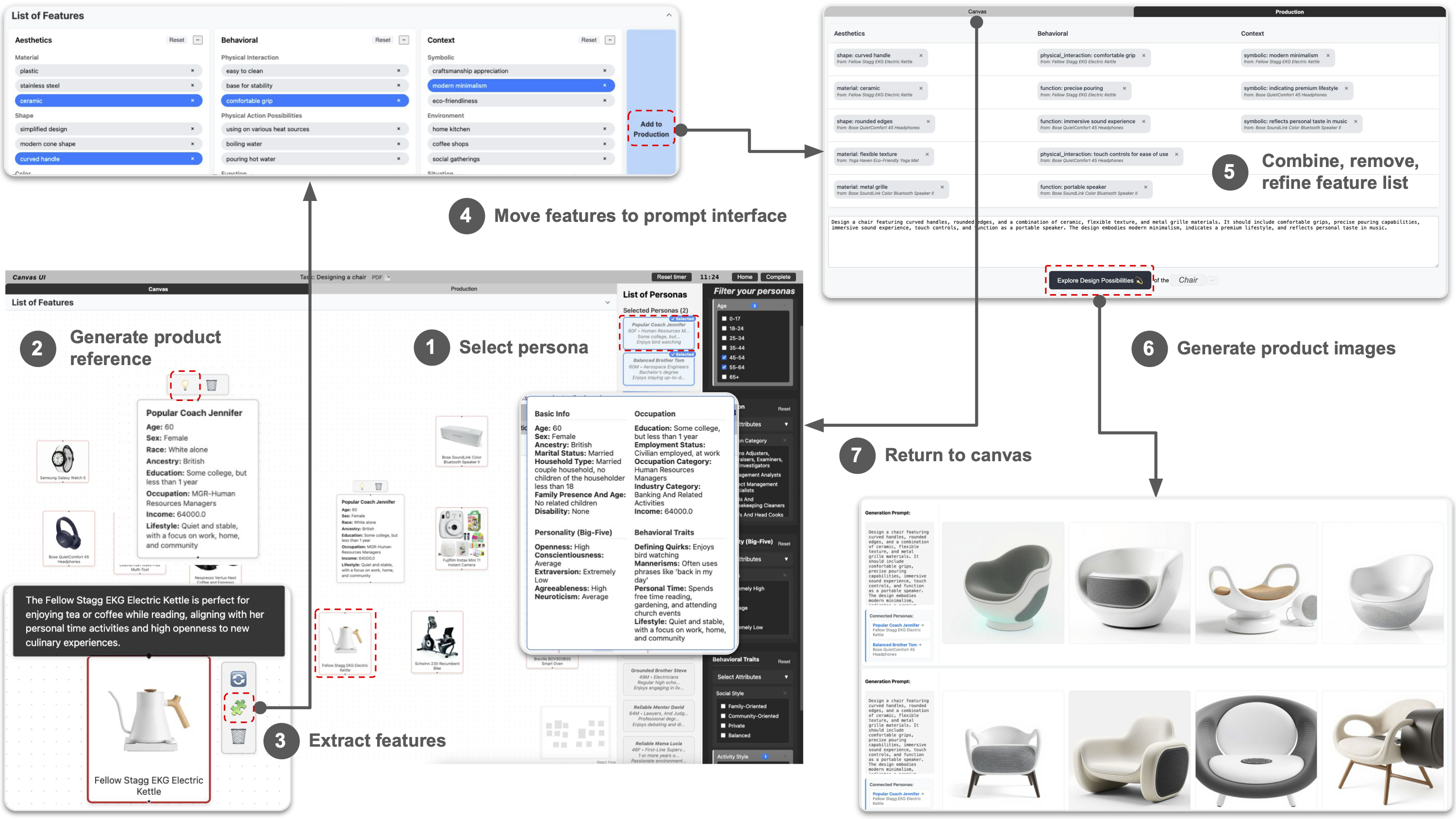}
  \caption{Iterative design process using Personagram. Upon filtering persona attributes, users can proceed to the ideation process by 1) selecting a persona, 2) generating product reference images, 3) extracting design features from product image, 4) identifying and moving features to the production UI (i.e., prompt interface), 5) refine feature list, 6) generate product images, iterate by remaining on the production UI or 7) returning to the canvas UI}
  \label{fig:interaction}
  \Description{A diagram illustrating the seven-step iterative design workflow within Personagram, connecting various interface components with directional arrows. The process flows as follows: 1) Select a user profile from the "List of Personas," 2) Generate "product reference" images (such as a kettle) linked to that persona, 3) Extract specific design features from the reference images, 4) Move these features to the "Production UI," 5) Combine, remove, and refine the feature list in the scaffolding interface, 6) Generate new product images (e.g., futuristic chairs) based on the refined features, and 7) Return to the "Canvas UI" to restart the cycle or iterate further.}
\end{figure*}

We instantiate our goals of scaffolding structured ideation workflows through the use of interactive buttons, each of which triggers a predefined prompt that incorporates details of the persona or persona-derived outputs. Users can click on the light bulb icon (\twemoji{bulb}) to generate 4-5 Product Reference Images (Fig. \ref{fig:teaser}D) at a time, which are displayed adjacent to the Persona Tile. Clicking on the product image summons a vertical tooltip bar, from which users can choose among a set of actions---image swap (rotating arrow icon), feature extraction (\twemoji{jigsaw}), and discard (\twemoji{wastebasket}). Clicking on the puzzle icon (\twemoji{jigsaw}) populates a feature list matrix (Fig. \ref{fig:interaction} (3)) above organized by aesthetic, behavioral and context dimensions, enabling user driven exploration of design concepts as they discover new product references and select desired features to use as building blocks for prototyping new product concepts (\textbf{DG2}).

\subsubsection{Production UI: Prototyping Support}
Once a user adds design features or switch to the production UI view (Fig \ref{fig:production}), three additional system components are displayed. The Dimensional Scaffolding interface (Fig \ref{fig:production}H) enables users to refine the feature list (organized by aesthetic, behavioral, and context dimensions) by removing previously selected features or adding more from the Canvas UI view (\textbf{DG2}). The Text-to-image Prompt Input box (Fig \ref{fig:production}I) updates dynamically based on changes made from the above feature list, but also allows direct modifications to the text prompt to better capture user intent (\textbf{DG3}). Finally, pressing the ``Explore Design Possibilities'' button sends a request to generate a new image based on the given prompt. The resulting images are displayed on the Image Gallery (Fig \ref{fig:production}J) in reverse chronological order with the input prompt and meta information about linked personas and products.

\subsection{Interaction Workflow}
Between the Canvas and Production UI, users can explore personas, explore concepts driven by personas, and prototype those concepts in an iterative fashion. The interaction workflow is illustrated in Fig \ref{fig:interaction}. The process begins in the Persona Exploration area, where users filter personas by demographic and behavioral attributes and preview detailed persona profiles. Once a persona is selected, it appears on the Canvas UI as a Persona Tile, where users can generate associated product reference images (\twemoji{bulb}). These reference images serve as visual inspiration, and users can interact with them directly by swapping images (\twemoji{wastebasket}), discarding irrelevant ones, or extracting design features (\twemoji{jigsaw}) such as shapes (\eg curved handle), physical interaction (\eg comfortable grip), and symbolic style (\eg modern minimalism)---organized by aesthetic, behavioral and context design dimensions. Extracted features can then be transferred to the Production UI to prepare for prototyping.

When users transition from exploration to prototyping, they move into the Production UI, which provides structured scaffolds for refining and translating selected features into concrete design outputs. Here, the Dimensional Scaffolding interface organizes features into aesthetic, behavioral, and contextual categories, giving users a structured lens for evaluating and editing their selections. These features automatically populate the text-to-image prompt input, allowing users to generate product images tailored to the chosen persona’s attributes and references. Generated outputs are displayed in the Image Gallery alongside the input prompts and linked metadata, enabling quick iteration, comparison, and revert back to previous prompt. Users can remain within the Production UI to refine prompts and continue prototyping or return to the Canvas UI to explore new personas and product references, thereby supporting a fluid, iterative workflow across ideation stages (\textbf{DG4}).


\begin{figure*}[t!]
  \includegraphics[width=\textwidth]{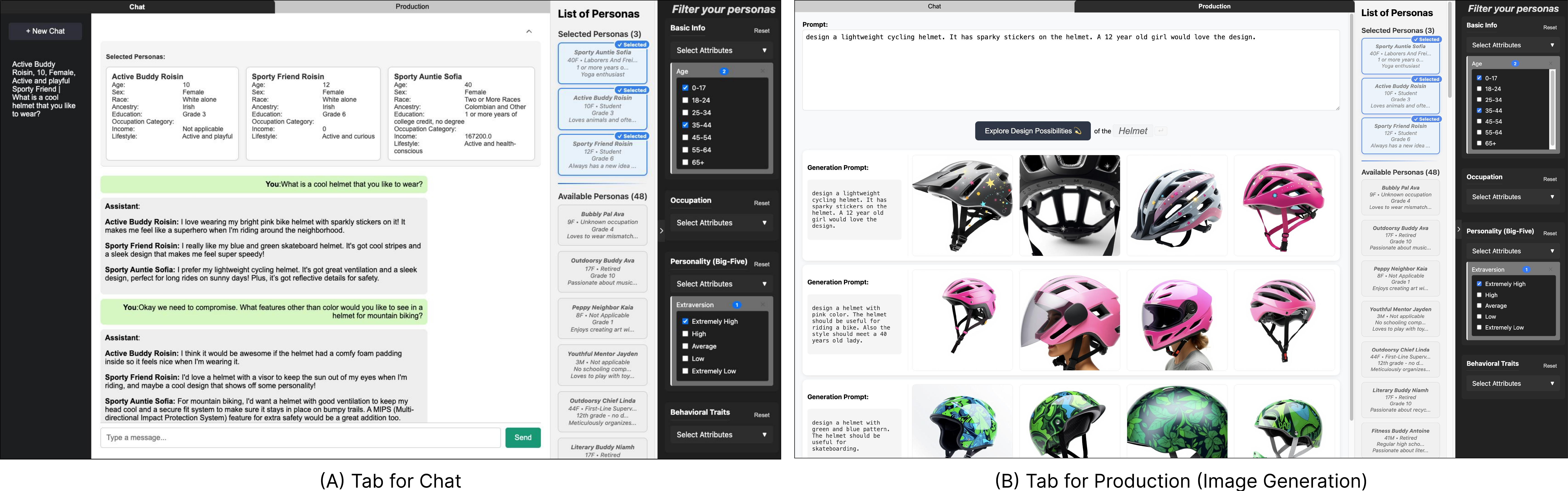}
  \caption{Baseline system. Users can interact with an LLM through a chat interface akin to commerical LLM products.}
  \label{fig:baseline}
  \Description{A side-by-side screenshot of the Baseline System interface, divided into two tabs: (A) Chat and (B) Production. The "Chat" tab (A) on the left features a conversational interface where the user interacts with three selected personas ("Active Buddy Roisin," "Sporty Friend Robin," and "Sporty Auntie Sofia") to discuss helmet designs in natural language. The "Production" tab (B) on the right shows a standard text-to-image workflow: a text prompt describing a "lightweight cycling helmet" is entered at the top, and the resulting AI-generated images of pink and green helmets are displayed in a grid below, alongside the list of selected personas and demographic filters.}
\end{figure*}

\subsection{Baseline Setup} 
\revised{We developed the baseline system to mimic how users interact with AI-powered personas in commercial chat-based systems, such as persona.design\footnote{\url{https://persona.design/}} and UXPressia\footnote{\url{https://uxpressia.com/ai-persona-chat}}. We retained the persona exploration interface components (\autoref{fig:teaser}A and B) from~\tool but replaced the Canvas UI (\autoref{fig:teaser}F) with a standard chat interface (see \autoref{fig:baseline} (A)). Clicking a persona enabled users to embed its detailed information into the chat context window. We provided users the freedom to interact with these personas via the prompt input box in flexible ways; for instance, users could choose to chat with the personas directly (role-play) or request the LLM to analyze the personas for brainstorming. Users could also manage multiple conversation threads with different sets of personas. For prototyping, the system provided a text prompt box for generating images through a T2I model comparable to commercial-grade T2I systems (see \autoref{fig:baseline} (B)). However, dimensional scaffolding was excluded because it was incompatible with the baseline use case.}

\subsection{Implementation Details}



We implemented~\tool as a web-based application using Next.js\footnote{\url{https://nextjs.org/}}. The system comprises two main components: the canvas (\textit{persona} and \textit{concept exploration}) and the production UI (\textit{prototyping}). To support persona exploration, we implemented hierarchical filters across four top-level categories: basic info, occupation, personality, and behavioral traits. Each category contains nested persona attributes (\eg~basic info > age, sex, marital status, disability, etc.). The detailed hierarchy is provided in Appendix~\ref{appx:attribute}. We built the canvas UI using ReactFlow\footnote{\url{https://reactflow.dev/}}. On this canvas,~\tool supports two key functions: persona-related product inference (\twemoji{bulb}) and feature extraction (\twemoji{jigsaw}) from inferred product images. For product inference, we designed chain-of-thought style~\cite{wei2022chain} reasoning prompts that instruct GPT-4o-mini to generate a list of persona-related products (see system prompts in~\autoref{appx:prompt}). Once~\tool receives the persona-relevant product list, it retrieves corresponding product images using the Google Image Search API. We chose GPT-4o-mini for its lightweight design, which enables responsive, real-time interactions. For the \textit{prototyping} process, we employed the Flux T2I model\footnote{\url{https://fal.ai/models/fal-ai/flux/schnell}} for its superior speed compared to other available image generation models. For both~\tool and the baseline implementation, we used the same AI models (\ie GPT-4o-mini and Flux).

\section{User Study}
To assess our hypotheses that designers would benefit from structured interactions with personas, we conducted a within-subject study with 12 professional designers.

\subsection{Participants}

\begin{table*}
\centering
\caption{Demographic and background of participants. *Participants' previous experiences with LLMs and text-to-image generation models; 1-never, 2-tried a few times, 3-monthly, 4-weekly, and 5-daily.}
\resizebox{\textwidth}{!}{%
\begin{tabular}{l l l l l l l l l l}
\toprule
    & & & & \multicolumn{2}{c}{Experience*} & \multicolumn{2}{c}{Session 1} & \multicolumn{2}{c}{Session 2}
    \\ \cmidrule(lr){5-6} \cmidrule(lr){7-8} \cmidrule(lr){9-10} 
    
    PID & Year & Occupation & Product & LLM & T2I & System & Task & System & Task\\
    \midrule
     P1 & 3 & Vehicle Designer & Camper, Trailer & 4 & 4 & Personagram & Chair & Baseline & Helmet \\ 
     P2 & 21 & Designer/Inventor & Toy & 2 & 1 & Personagram & Chair & Baseline & Helmet \\ 
     P3 & 20 & Design Director & Furniture, Faucet & 4 & 3 & Personagram & Chair & Baseline & Helmet\\ 
     P4 & 4 & Industrial Designer & Footwear & 4 & 2 & Personagram & Helmet & Baseline & Chair\\ 
     P5 & 15 & Industrial Designer & Helmet, Vehicle & 2 & 4 & Personagram & Helmet & Baseline & Chair\\ 
     P6 & 2 & Industrial Designer & Planter & 4 & 3 & Personagram & Helmet & Baseline & Chair\\ 
     P7 & 15 & Product Designer & Jewerly, Vehicle & 5 & 4 & Baseline & Chair & Personagram & Helmet\\ 
     P8 & 15 & Product Engineer & VR device & 5 & 3 & Baseline & Chair & Personagram & Helmet\\
     P9 & 5  & Industrial Designer & Radio & 4 & 1 & Baseline & Chair & Personagram & Helmet\\
    P10 & 25 & Product Designer & Gadget, Vehicle & 5 & 4 & Baseline & Helmet & Personagram & Chair\\
    P11 & 25 & Spatial Designer & Museum exhibit & 4 & 4 & Baseline & Helmet & Personagram & Chair\\
    P12 & 2 & Industrial Designer & Electronic gadget & 5 & 5 & Baseline & Helmet & Personagram & Chair\\
    
\bottomrule
\end{tabular}
}
\label{tab:participants}
\end{table*}

We recruited 12 professional physical product designers through Upwork\footnote{\url{https://www.upwork.com/}} and internal mailing lists. Eligible participants (18 years or older, English-speaking, professional physical product designers) represented a broad range of expertise in industrial design, from 2 to 25 years. Their design domains varied horizontally, including areas such as vehicle, toy, furniture, helmet, planter, electronic gadget, and virtual reality (VR) device. Three participants (P1, P10, and P12) held Master's degrees, while the rest held bachelor's degrees. We screened out designers primarily working on non-physical products (\eg music, games, UI/UX). We also excluded design students (\eg Master in Design) who lack professional experience. We asked participants about their prior experiences with generative AI, including both LLMs (\eg ChatGPT, Claude, Gemini) and T2I tools (\eg Midjourney, DALL·E). All participants had used an LLM, while two (P2 and P9) had never used any image generation tools. We tabulated demographic details in~\autoref{tab:participants}.

\begin{figure*}
  \includegraphics[width=0.9\textwidth]{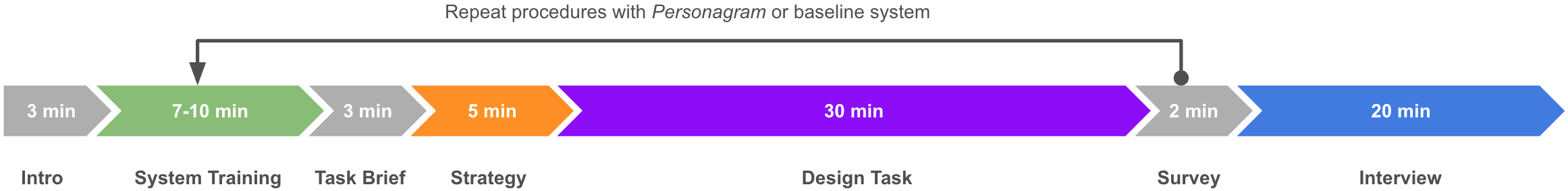}
  \caption{Study procedure flow.}
  \label{fig:procedure}
  \Description{A timeline diagram depicting the flow of the user study session. The procedure begins with an "Intro" (3 min). A recurring block of five phases follows: "System Training" (7–10 min), "Task Brief" (3 min), "Strategy" (5 min), "Design Task" (30 min), and "Survey" (2 min). An arrow connects the end of the Survey phase back to the start of System Training, indicating that this block is repeated for the second condition (using either Personagram or the Baseline system). The session concludes with a final "Interview" (20 min).}
\end{figure*}

All eligible participants who passed the screener provided informed consent and joined a 120-minute Zoom session. Each session included an introduction, system training, review of the task brief, a short strategy discussion, the design task, a post-task survey, and an interview (\autoref{fig:procedure}).

\subsubsection{System Training and Task Brief}
Participants first completed a five-minute tutorial followed by free-form exploration to familiarize themselves with key system features. Once comfortable, they read a pdf of the task brief for approximately three minutes. The brief described two design tasks: a chair for school-aged children and a helmet for older adults. 

\subsubsection{Design Task}
Before starting the design task, participants outlined a short design strategy, hypothesizing about personas and attributes that might guide their ideation. This step was added based on pilot studies, where participants reported feeling overwhelmed by the number of available personas and filters. During the task, participants were encouraged to think aloud, and facilitators provided clarification only when necessary. Participants were asked to select two favorite generated images at the end of the task. We used a 2x2 counterbalancing scheme to vary the order of tool (\tool or Baseline) and design task (chair vs. helmet) across participants.

\subsection{Procedure}

\subsubsection{Post-study Survey \& Interview}
After completing the design task, participants rated their own designs and evaluated their experience with each system using the measures described above. Finally, we conducted a 20-minute semi-structured interview to capture participants’ attitudes, reflections, and overall experiences with the tools. The interview questions included, but were not limited to, topics about perceived differences across the two systems, how personas affected their decision making, and how designers envision using similar systems in their professional practice and workflow. Upon completion of the study, we provided \$100 as compensation via the Upwork platform.

\subsection{Measures}
\revised{We examined participants' behavioral patterns across the three processes (persona exploration, concept exploration, and prototyping) by recording usage logs from both systems. The system recorded timestamps, feature interaction counts, prompt queries, generated images, session audio, and screen activity.}
\revised{We used these interaction logs to quantify user engagement, specifically by counting the total number of button clicks (\twemoji{bulb}: bulb, \twemoji{heavy_plus_sign}: plus, and \twemoji{jigsaw}: puzzle) in~\tool and the number of submitted text queries in the baseline.} 
To further assess perceived user experience, we employed the Creativity Support Index (CSI)~\cite{cherry2014quantifying}, NASA-TLX~\cite{hart1988development}, and self-perceived experience with GenAI models~\cite{wu2022ai}, comparing \tool against the baseline system. 

In addition, we measured perceived performance of \tool across persona exploration, concept exploration, and prototyping processes with a 7-point Likert scale (\autoref{fig:survey_personagram}). The questions included: 1) [\textbf{organize personas}] It was easy to organize personas by different dimensions (e.g., job, age, etc.); 2) [\textbf{guide exploration}] The persona previews contained enough information to guide exploration; 3) [\textbf{discover inspirations}] It was easy to discover new product inspirations; 4) [\textbf{influence thinking}] The product inspirations helped me think in new directions; 5) [\textbf{bridge personas}] The connection between persona and product features was clear to me; 6) [\textbf{design intentionally}] I made a conscious effort to design for the persona in mind; 7) [\textbf{map outcomes}] The design consequences of considering different personas were clear to me.


\subsection{Analysis}
\revised{For quantitative measures (survey responses), we ran Wilcoxon Signed-Rank test to compare results between conditions due to the ordinal nature of Likert scale in these surveys~\cite{choi2024creativeconnect}. Given the sample size ($N=12$), we argue that a strictly dichotomous interpretation of results—categorizing them simply as significant or non-significant—obscures the uncertainty and strength of the outcomes~\cite{rafi2020}. To provide a more transparent and informative analysis, we report 95\% compatibility intervals (CI), allowing for an assessment of the magnitude and precision of the user feedback.} For analyzing screen recordings, three of authors independently conducted manual coding across the three phases (persona exploration, concept exploration, and prototyping), transitions between them, and total duration of each phases following the procedure outlined in~\cite{zhou2025productmeta}. For interview responses, once we retrieved auto-transcribed scripts of all participants, we conducted thematic analysis~\cite{virginia2006using} assisted by GPT-5\footnote{\url{https://openai.com/gpt-5/}}, following the protocol in~\cite{Goyanes2025}. After preparing the transcripts, GPT-5 was first instructed to identify overarching themes across the dataset. We then iteratively refined these themes by prompting for convergences, divergences, representative quotes, and theme-specific summaries. Two of the authors reviewed these outputs, cross-checking against the transcripts and refining theme boundaries. Finally, they reviewed and validated the themes to reach consensus, ensuring that all findings were grounded in the raw transcripts.

\section{Results}
\begin{figure*}
  \includegraphics[width=\textwidth]{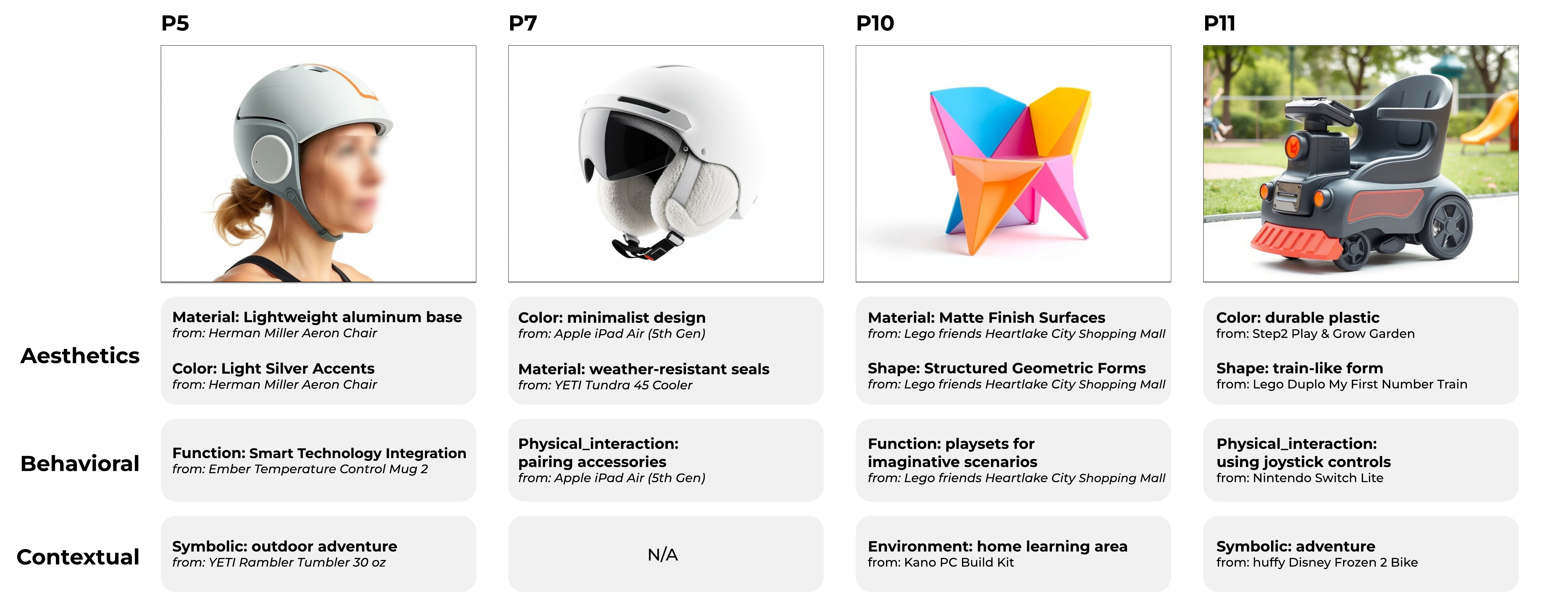}
  \caption{Example images of outcomes generated by participants using~\tool (two helmets for older adults, and two chairs for children). The figure also shows the corresponding features participants selected for incorporation into prototyping, extracted from persona-relevant products in~\tool.}
  \label{fig:outcomes}
  \Description{Figure 9 displays a gallery of four final design concepts generated by participants (P5, P7, P10, and P11), arranged in columns. The top row shows the AI-generated images: two modern helmets designed for older adults (P5, P7) and two playful chairs designed for children (P10, P11). Below each image is a list of the specific attributes the participant extracted from other products to create the design, categorized by Aesthetics, Behavioral, and Contextual dimensions. P5 (Helmet): A sleek grey helmet incorporating "Lightweight aluminum" from a Herman Miller Chair and "Smart Technology" from an Ember Mug. P7 (Helmet): A white minimalist helmet incorporating "weather-resistant seals" from a YETI Cooler and "pairing accessories" from an iPad Air. P10 (Chair): A multi-colored, geometric chair incorporating "Structured Geometric Forms" and "Matte Finish Surfaces" from a Lego Friends set. P11 (Chair): A train-shaped ride-on chair incorporating a "train-like form" from Lego Duplo and "joystick controls" from a Nintendo Switch.}
\end{figure*}

All participants (except P2) expressed strong preferences for using~\tool. Some (\eg~P6 and P12) requested extended public access so they could continue applying the system to their own projects.~\autoref{fig:outcomes} presents four example outcomes (two helmets and two chairs) that participants marked as favorites. Below, we first summarize the user logs of~\tool and the baseline. We then present our findings for each research question: impact on the creative process (RQ1), perceived transparency and reliability (RQ2), and perceived value of the tool (RQ3).

\subsection{\revised{User Engagement with Personas through Buttons vs. Prompt}}

\revised{Usage logs reveal that \tool's structured interface fostered higher engagement with personas compared to the open-ended baseline ($W = 55$, $p = 0.002$), despite apparent constraints on user freedom (see~\autoref{fig:logs}; raw data in Appendix~\autoref{tab:userlogs}). Participants clicked interaction buttons (\twemoji{bulb}, \twemoji{heavy_plus_sign}, and \twemoji{jigsaw}) 187 times in~\tool, whereas they submitted only 92 text queries in the baseline. This increased amount of interaction was driven by the explicit affordances of the interface, particularly the \twemoji{jigsaw} button, which alone accounted for 83 interactions. By enabling the decomposition of persona-derived product images with a single click,~\tool lowered the interaction cost for a task that was theoretically possible in the baseline but rarely performed without such scaffolding. Consequently, users utilized the buttons to explore personas more frequently than they were willing to type out equivalent text queries.}

\begin{figure*}
  \includegraphics[width=0.9\textwidth]{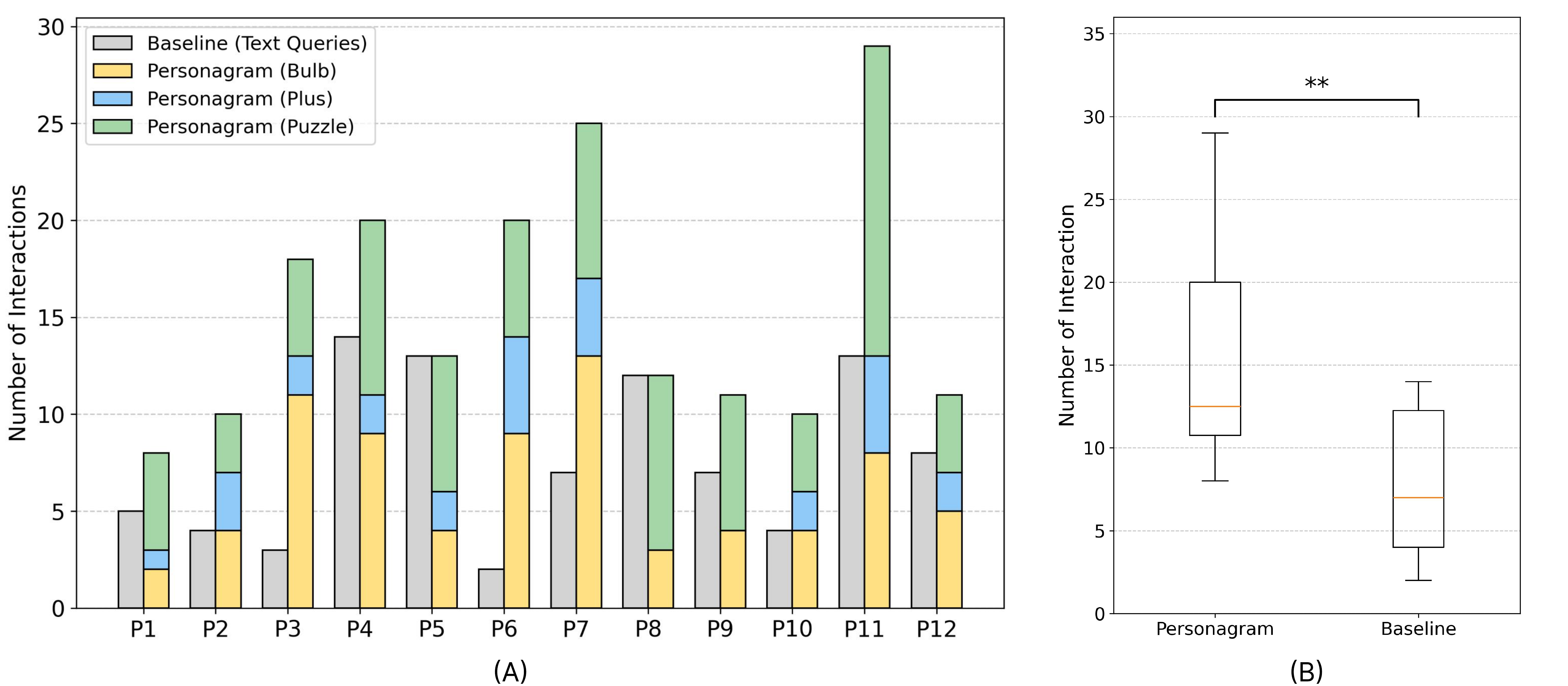}
  \caption{\revised{Usage logs showing (A) the number of interactions for each participant using~\tool versus the baseline, and (B) a box plot comparing the total interactions between conditions. In (A), the stacked bars for~\tool break down the usage of three features (\twemoji{bulb}: bulb, \twemoji{heavy_plus_sign}: plus, and \twemoji{jigsaw}: puzzle), while baseline bars represent the number of text queries. ($**$: $p < .01$).}}
  \label{fig:logs}
  \Description{Figure 6 presents two charts comparing user interactions between the Personagram and Baseline systems.(A) A bar chart showing the "Number of Interactions" for 12 individual participants (P1–P12). For each participant, a grey bar represents Baseline usage ("Text Queries"). A second, taller stacked bar represents Personagram usage, broken down into three specific actions: "Bulb" (yellow), "Plus" (blue), and "Puzzle" (green). In almost all cases (e.g., P3, P4, P6, P7, P12), the Personagram bar is significantly higher than the Baseline bar.(B) A box plot summarizing the total interactions for all participants. The "Personagram" box is higher and taller (median $\approx$ 13, range 8–29) compared to the "Baseline" box (median $\approx$ 7, range 2–14). A bracket linking the two bears a double asterisk (**), indicating a statistically significant difference with $p < .01$.}
\end{figure*}


\revised{By contrast, with the open-ended nature of the baseline, participants used prompt-based chat to interact with personas, anthropomorphizing them while performing analysis, brainstorming concepts, and generating text-to-image prompts. Participants treated personas as focus group members, interviewing them individually to uncover preferences (\eg favorite colors, characters), behavioral traits (\eg personas' posture in class), or design requirements (\eg "what kind of chair do you want?"). While this allowed for flexible inquiry, the cognitive effort required to formulate these questions likely contributed to the lower overall volume of interaction (see Appendix~\ref{appx:analysis} for query categories).}

\revised{Overall, the results highlight the value of structured scaffolding. While~\tool replaced the open-ended prompt interaction with three defined buttons, this design choice operationalized complex interactions into low-friction actions. Conversely, while the baseline offered theoretically unlimited freedom, the lack of explicit scaffolding resulted in lower engagement volume. This interface distinction influenced the creative process, perceived transparency, and the perceived value of the tool, as we detail in the following sections.}

\begin{figure}
  \includegraphics[width=0.46\textwidth]{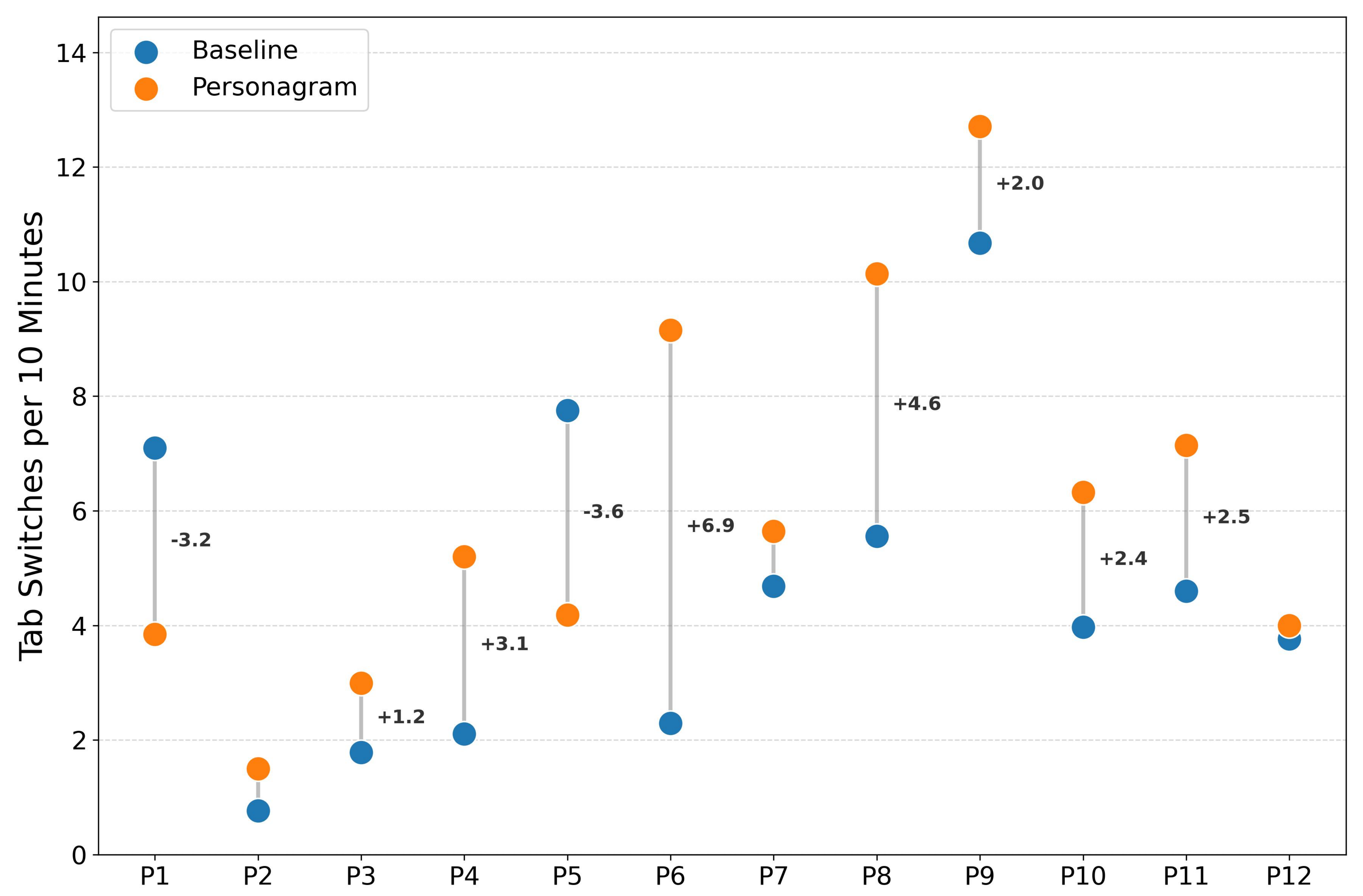}
  \caption{\revised{Comparison of tab switching rates per 10 minutes. Except for P1 and P5, all participants demonstrated a higher switching rate with~\tool. This increased activity implies more frequent reviews of personas and related concepts compared to the baseline.}}
  \Description{Dot plot comparing tab-switching rates per 10 minutes for twelve participants under two conditions: Baseline and Personagram. Each participant has two points connected by a vertical line, showing the change between conditions. For most participants, the Personagram condition shows higher tab-switching rates than the Baseline, with increases annotated next to the connecting lines. Two participants, P1 and P5, show lower switching rates with Personagram. Overall, the plot indicates increased tab-switching activity when using Personagram.}
  \label{fig:tabswitch}
\end{figure}

\subsection{Impact on Creative Process (RQ1)}



\subsubsection{Frequent Idea Exploration with Reduced Burden in Prompting}

\begin{figure*}
  \includegraphics[width=0.9\textwidth]{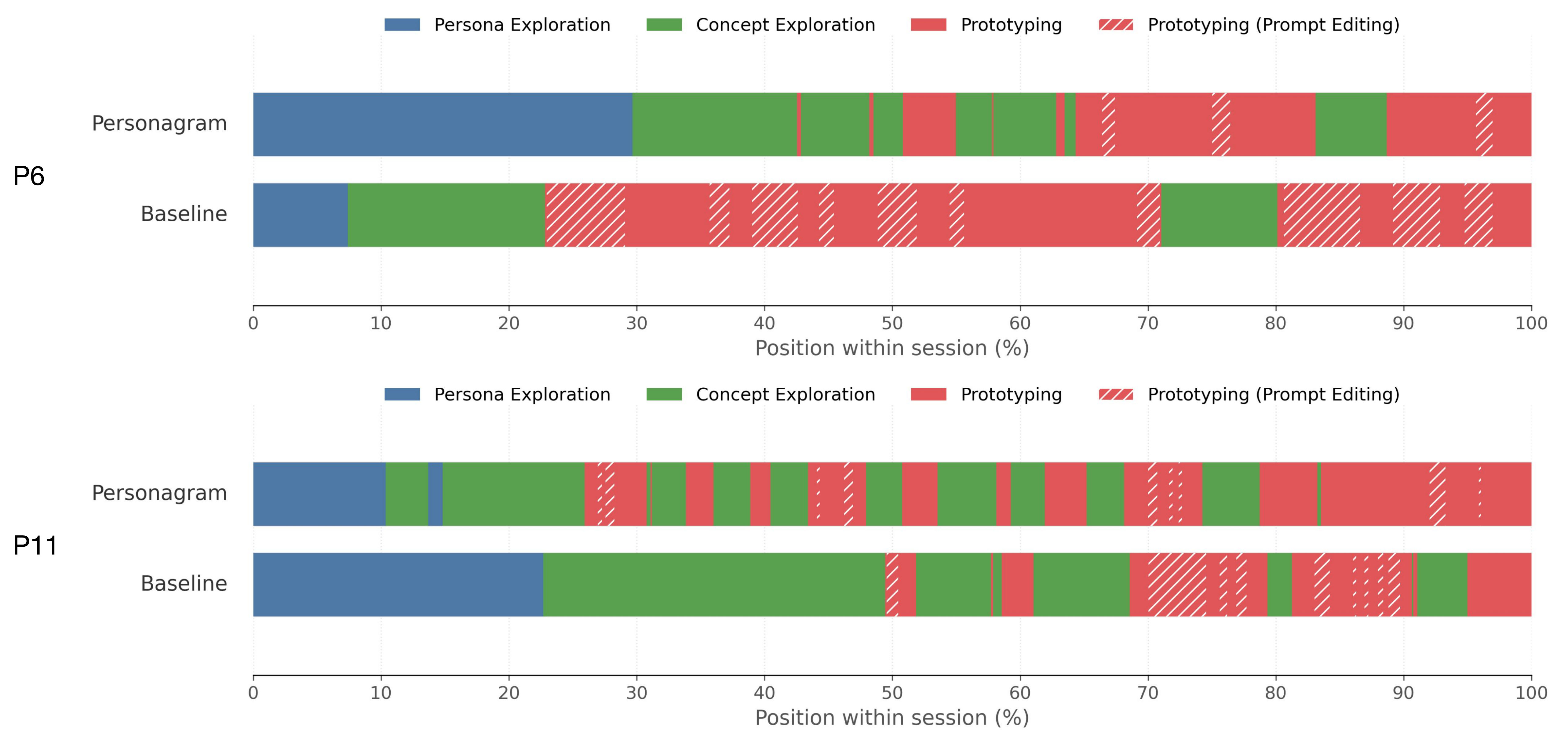}
  \caption{\revised{Normalized session timelines for P6 and P11. When using~\tool, both participants switched between exploration and prototyping tasks more frequently compared to the baseline. In contrast, the baseline sessions were characterized by longer, uninterrupted blocks of prototyping (notably in P6).}}  
  \label{fig:timeframe}
  \Description{Figure 8 illustrates the normalized session timelines for two participants, P6 and P11, comparing their activity flows in the Personagram condition versus the Baseline. The charts use color-coded bars to represent four activity types: "Persona Exploration" (blue), "Concept Exploration" (green), "Prototyping" (solid red), and "Prototyping - Prompt Editing" (hatched red). The x-axis represents the "Position within session (\%)" from 0 to 100. For P6: The Baseline timeline shows a linear progression: a brief exploration phase followed by a massive, uninterrupted block of "Prototyping" and "Prompt Editing" spanning roughly 75\% of the session. In contrast, the Personagram timeline is highly fragmented, showing frequent alternations between green "Concept Exploration" and red "Prototyping" blocks throughout the session. For P11: A similar pattern appears. The Baseline session is characterized by long, distinct phases (a large block of exploration followed by a large block of prototyping). The Personagram session shows a more interleaved pattern, where the participant repeatedly switches back and forth between exploring concepts (green) and prototyping (red). Overall, the visual data indicates that Personagram encourages a non-linear, iterative workflow, whereas the Baseline promotes a more sequential, "waterfall" style of working.}
\end{figure*}

%
With~\tool, participants explored diverse ideas more frequently compared to the baseline. Despite its structured UI workflow, Creativity Support Index (CSI ratings were comparable to the baseline condition (see~\autoref{tab:index}), indicating that the structured interface did not compromise participants' sense of creative freedom. User logs confirm this pattern: participants switched tabs between \textit{exploration} (canvas or chat) and \textit{prototyping} (production) about 30\% more often in~\tool (6.0 vs. 4.6 per 10 mins; see~\autoref{tab:userlogs}). For example, P6 switched tabs between \textit{prototyping} (production) and canvas 13 times in~\tool, compared to only 3 times in the baseline (see~\autoref{fig:timeframe}). This behavior reflects tighter exploration–prototyping loops, supporting faster and more iterative design processes. Participants stated:

\begin{quote}
P9: ``It allowed me to iterate through ideas faster without worrying about wording.''
\end{quote}

While participants iterated more with~\tool, task session times were slightly shorter. Although participants explored design inspirations more frequently, they spent less time editing prompts. For 8 of 12 participants, baseline required longer prompt-editing. This indicates a shift: in~\tool, users were able to focus less on `writing better prompts' and more on exploring persona-related inspirations (\ie, product references and extracted features) on the canvas, as well as evaluating generated outputs on the production tab. The reduced prompt-editing time appears to offset the cost of additional exploration, resulting in shorter sessions to complete the task.

\subsubsection{Satisfactory Outcomes through Fine-Grained Inspirations}

Despite spending less overall task time with~\tool, participants reported producing more satisfactory outcomes. Shorter sessions were made possible by reducing the effort required for prompt editing, which allowed participants to allocate more attention to exploring inspirations (\ie~product images and extracted features) and evaluating generated outputs.


\begin{quote}
P9: ``It was easier to visualize where I was going, instead of doing all of the thinking and working in words, and then getting a picture at the end that may or may not have anything to do with what I had in my head.''
\end{quote}
\begin{quote}
P10: ``By breaking it [product images] into discrete qualities, it [\tool] makes it a lot easier to manipulate, rather than thinking of text to type.''
\end{quote}

\revised{This efficiency did not come at the expense of quality. On the contrary,~\tool consistently supported higher perceived value. For the CSI dimension \texttt{Results Worth Effort}, participants rated \tool higher than the baseline ($6.42$ vs. $5.25$). The analysis shows an estimated benefit of 1.00 point, with data compatible with a true difference ranging from 0.00 to 2.50 (95\% CI). While the lower bound is close to zero, the interval extends to a substantial magnitude, suggesting a practical benefit. Furthermore, the NASA-TLX dimension of \texttt{Performance} showed a benefit, with data compatible with an estimated improvement of 0.00 to 1.50 points favoring \tool (see~\autoref{tab:index}). These results suggest that the structured, fine-grained inspirations provided by \tool not only accelerated the design process but also supported participants in achieving outcomes they perceived as worth their effort.}

\subsection{Perceived Transparency and Reliability of Responses (RQ2)}

\subsubsection{Improved Transparency Through Multimodal Workflow}

\begin{figure*}
  \includegraphics[width=0.8\textwidth]{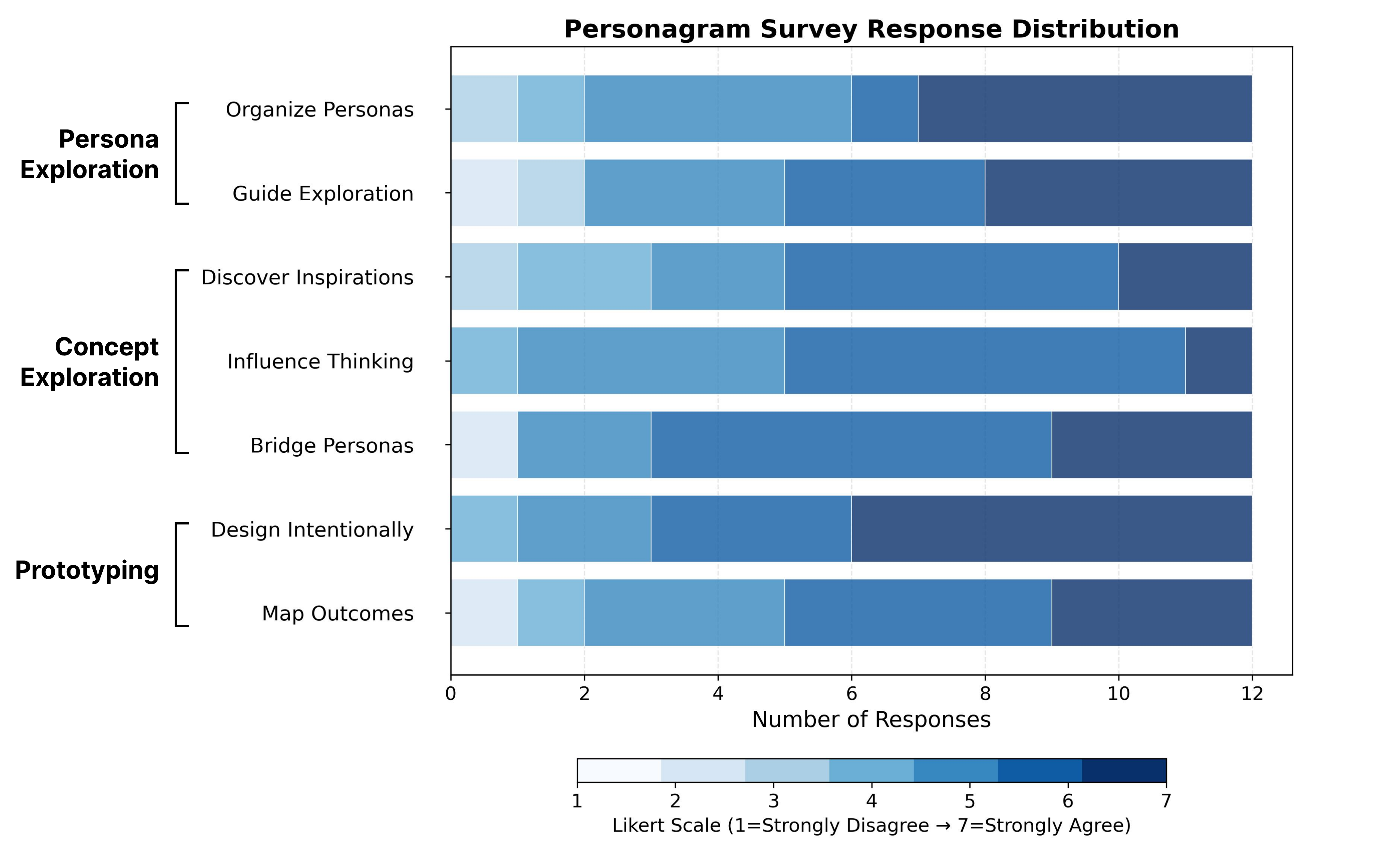}
  \caption{Stacked number of responses for survey questions about Personagram features.}
  \label{fig:survey_personagram}
  \Description{Figure 10 is a horizontal stacked bar chart titled "Personagram Survey Response Distribution." It visualizes Likert scale ratings (from 1="Strongly Disagree" in light blue to 7="Strongly Agree" in dark blue) given by 12 participants across three system categories: Persona Exploration: Includes the items "Organize Personas" and "Guide Exploration." Both show strong consensus, with the majority of responses in the dark blue (6–7) range. Concept Exploration: Includes "Discover Inspirations," "Influence Thinking," and "Bridge Personas." Responses are predominantly positive, though "Influence Thinking" shows slightly more variance with some mid-range scores. Prototyping: Includes "Design Intentionally" and "Map Outcomes." "Design Intentionally" shows particularly high agreement, with nearly all participants rating it a 5 or above. Overall, the chart demonstrates that participants perceived the system's features as effective for supporting specific design activities, with a heavy skew toward "Agree" and "Strongly Agree."}
\end{figure*}

\revised{Participants (\eg~P1, P5, P6, P9, P10, P12) reported that \tool provided a more transparent and interpretable experience compared to the baseline. Analysis of the self-perceived experience features~\cite{wu2022ai} supports this, showing a benefit for \tool regarding transparency. The analysis shows an estimated benefit of 1.50 points for \tool ($5.33$ vs. $3.92$). The data are compatible with a true difference ranging from 0.00 to 2.50 (95\% CI). This compatibility interval indicates that while the magnitude of the benefit is uncertain due to the small sample size ($N=12$), the direction of the effect favors \tool. In addition, most participants (9 out of 12) selected \textit{agree} or \textit{strongly agree} for the transparency-related item: \texttt{Bridge Personas}---``the connection between persona and product features was clear to me'' (see~\autoref{fig:survey_personagram}).}

Participants (P5, P6, P9, P10, P12) attributed this transparency to the multimodal pipeline of \tool, which makes it easier to connect personas with final outcome features. The process transitions across modalities: starting from persona descriptions (text), which are translated into product visual references (images), then decomposed into features and prompts (text), and finally used to generate outcome images (image). This structured cross-modal flow helps participants interpret what specific keywords in the prompt mean and trace which product images the features originated from. As a result, outcomes were perceived as more interpretable, since the process established a transparent link between personas, prompts, and generated images.

\begin{quote}
P5: ``If I'm making a helmet for an outdoorsy person. I think usually the first question I'm asking myself is, what is it used for? So, the context is very important. So, engaging in outdoor activities, Weekend getaway, attending a festival, used in nature. This is really good, yeah, symbolic, this is really nice. I like this sense of exploration. Usually I like to think more abstractly, and [...] not too literal on the aesthetics until you get to the core of the symbolic nature of the design.''
\end{quote}

\begin{table*}
\centering
\caption{The results of three survey responses: creativity support index (CSI), perceived taskload (NASA-TLX), and self-perceived experience on AI-powered system features. ($+$: $p < .1$, *: $p < .05$)}
\begin{tabular}{r r c c c c c c c c c}
\toprule
    & & \multicolumn{2}{c}{Personagram} & \multicolumn{2}{c}{Baseline} & \multicolumn{3}{c}{Statistics}
    \\ \cmidrule(lr){3-4} \cmidrule(lr){5-6} \cmidrule(lr){7-9} 
    
     &  & mean & std & mean & std & $W$ & $p$ & Sig. & MedianDiff. (95\% CI) \\
    \midrule
    \multirow{6}{*}{CSI}
      & Enjoyment & 5.58 & 1.38 & 5.25 & 1.42 & 13.0 & 0.480 &  \\ 
      & Exploration & 5.0 & 1.81 & 5.08 & 1.51 & 10.0 & 0.915 &  \\ 
      & Expressiveness & 5.42 & 1.31 & 4.58 & 1.24 & 14.5 & 0.180 &  \\ 
      & Immersion & 5.17 & 1.34 & 4.67 & 1.44 & 12.0 & 0.394 &  \\ 
      & \textbf{Results Worth Effort} & 6.42 & 0.79 & 5.25 & 1.66 & 7.0 & 0.057 & $+$ & $+1.00$ [0.00, 2.50]\\ 
      \midrule
     \multirow{6}{*}{NASA-TLX}
      & Mental & 3.83 & 1.12 & 3.25 & 1.14 & 22.0 & 0.215 &  \\ 
      & Physical & 1.83 & 0.84 & 1.67 & 0.78 & 10.5 & 0.317 &  \\
      & Temporal & 2.92 & 1.00 & 3.00 & 1.41 & 32.5 & 0.776 &  \\      
      & Effort & 2.75 & 1.29 & 3.25 & 1.54 & 12.0 & 0.201 &  \\
      & \textbf{Performance} & 5.67 & 0.49 & 5.00 & 1.13 & 0.0 & 0.027 & $*$ & $+0.50$ [0.00, 1.50] \\                  
      & Frustration & 1.58 & 1.00 & 2.00 & 1.04 & 22.5 & 0.458 &  \\
      \midrule
      & Match goal & 5.75 & 0.97 & 5.17 & 1.27 & 20.0 & 0.162 & \\ 
     Self-perceived & Think through & 5.00 & 1.81 & 5.17 & 1.64 & 36.0 & 1.000 & \\ 
     experience on   & \textbf{Transparent} & 5.33 & 0.98 & 3.92 & 2.07 & 10.5 & 0.026 & $*$ & $+1.50$ [0.00, 2.50]\\ 
     AI system & Controllable & 5.17 & 1.27 & 4.67 & 1.23 & 15.0 & 0.121 \\ 
       & Collaborative & 5.25 & 1.66 & 4.83 & 1.80 & 25.5 & 0.402 & \\
\bottomrule
\end{tabular}
\label{tab:index}
\end{table*}

Here, P5 emphasized that transparency was not only about seeing the connection between features and outcomes, but also about enabling designers to reason abstractly about symbolic meaning. By surfacing contextual cues (\eg outdoor activities, festivals, nature), \tool supported participants in interpreting personas at a higher, conceptual level, rather than only through literal aesthetic translation.

\begin{quote}
P6: ``Absolutely, ... these [keywords extracted from the visual references] helps you visualize in the prompt. [In \tool], you physically take these product features [from product visuals] on canvas and applies them here [production prompt]. \textbf{Seeing} what they would have makes more sense [in \tool]. But just \textbf{hearing} about what they like, you don't really see from the design perspective what they look like [in the baseline]. For example, you don't really know what they mean by `sturdy contruction' [without seeing visuals] because that can take a lot of different forms.
\end{quote}



However, this transparency does not come for free. Participants (\eg~P3, P5, P6, P8, P9) noted that the multimodal workflow of~\tool nudges user input. Specifically, they had to: (1) read text (persona descriptions), (2) review images (persona-related product), (3) read and select texts (extracted features from those images), and (4) generate and review outcome images. This multimodal transition process would demand more mental effort compared to the free-form LLM baseline, where users could simply ask questions without going through multiple steps. 


\begin{quote}
P10: ``It's been super helpful to be able to chunk the cognitive units into discrete blocks, so I can add and remove, and I can \textbf{see} how that affects things [in \tool], rather than \textbf{the vagary of how the prompt is worded} [in the baseline].''
\end{quote}


\revised{Consistent with these observations, NASA-TLX results suggested directional trends, with participants reporting slightly higher \texttt{Mental Effort} (3.83 vs. 3.25) and lower \texttt{Frustration} (1.58 vs. 2.00) when using \tool, though these differences were not statistically significant (see~\autoref{tab:index}).} This trade-off highlights that while~\tool required additional user investment, it also yielded meaningfully more satisfactory and transparent outcomes (\texttt{Results Worth Effort} and \texttt{Performance} in~\autoref{tab:index}). In short, \tool enables users to infuse their input and intent into the process, resulting in experiences that are both more transparent.

\subsubsection{Perceived Reliability Through Structured Workflow}


Our analysis of interview data revealed that many participants (\eg, P3, P4, P5, P8, P10, P12) reported stronger trust and confidence when using~\tool. Participants emphasized that the inferred product lists and extracted features felt plausible and objectively true, since they could be traced back to actual personas and products. Unlike free-form chat, our system's structured workflow constrained outputs to recognizable design cues, which reduced the sense of randomness. As a result, participants perceived the concept exploration process in~\tool as more valid and reliable, with little doubt about the system responses. 

\begin{quote}
P4: ``It [\tool] seemed limited, but in a way that I could trust more. Pulling these products that these people are liking, and it’s telling me objectively true things about the product, like, it's made out of metal, or it’s sleek. It seemed true. It seemed trustworthy.''
\end{quote}

Some participants, such as P4 and P8 also appreciated the ability to talk to the personas to inquire specific details about them, including lifestyle, specific contextual experiences (e.g., school), preferences, among others. 

\begin{quote}
P4: ``They had the persona Olivia, who is a 16-year-old girl who is in school, and, you know, she's a little neurotic. I was able to kind of ask her, questions that felt [...] You know, a teenage girl has a very specific lifestyle, has a very specific day-to-day [...] I was able to get more in-depth answers about specific things.''
\end{quote}

In contrast, however, participants doubted the reliability of the baseline with chatting interface, questioning the validity of its responses. The free-form nature of the baseline allowed users to ask anything. However, since \textit{any input} is possible, this also meant that \textit{any output} could be generated in return. This unstructured nature could increase the likelihood of unreliable responses~\cite{kalai2025languagemodelshallucinate}.


\begin{quote}
P8: ``[baseline] offers super specific niche products. It doesn't mean that all 8-year-old males like this product. Just giving you some random things that doesn't really help to design it. I mean, firstly, it's unreliable, and secondly, if you ask 10 times, they'll give you 10 different responses.''
\end{quote}

\tool is not free-form; it introduces constraints by grounding responses in product images as design references. Participants found this approach highly reliable and plausible, with little need to question the validity of the outputs. Moreover, because each step in the pipeline performed a clear, well-defined task, participants reported greater trust in the results and ultimately felt more confident in their outcomes (P10). This distinction highlights an intriguing dynamic: while constraints limit flexibility, they also foster stronger perceptions of reliability and trust.






\subsection{Perceived Value of Personagram in Professional Practice (RQ3)}

\subsubsection{Lowering Barriers to Act on Personas}
Participants described \tool as reducing traditional barriers to engaging with personas by making them faster and easier to generate. Except for one participant who expressed limited appreciation of persona support, most participants (P1, P4--12) emphasized its potential for significant time and cost savings. As P11 noted, 
\begin{quote}
P11: ``You could imagine a big company spending a few weeks just trying to get those personas together, and if it's an agency, they would charge a company like, like, way too much money to do that.''
\end{quote}
P5 similarly contrasted conventional workflows with the efficiency of \tool:
\begin{quote}
P5: ``Normally it takes maybe a week of work to come up with personas… here you can go directly from personas to visual inspirations.''
\end{quote}

A second barrier came from organizational structures in which persona research is outsourced to other teams, leaving designers with little direct connection to underlying data. Several participants reported that this disconnect limited their ability to use personas effectively in their own design work. As P12 explained,
\begin{quote}
P12: ``I receive the persona from the research team [...] Normally, like, the leader won't tell you what he prefers. He gives you direction, and you give lots of concept, and he'll say, `Oh, I like that.' But he never asks why. But as a designer, I have to make sure that [I] won't be at fault.''
\end{quote}
Here, P12 emphasized the responsibility designers feel to produce defensible concepts despite limited guidance. He noted that \tool could reduce the time otherwise spent reconstructing or validating persona insights, freeing more time for ideation.


Beyond efficiency and organizational hurdles, participants highlighted how \tool could expand ideation by surfacing unexpected directions and helping uncover blind spots. P1 appreciated how persona-linked product prompts sparked ideas (e.g., Pokemon cards) outside of their initial expectations. Others, like P12, found the system especially valuable for uncovering overlooked user pain points. Initially unsure how to approach designing for children, P12 found the extracted keywords and product inference features particularly useful:
\begin{quote}
P12: ``I didn't figure out the painpoint [of children] in the first place, but the persona [in Personagram], it's really helped me find out what is the daily life that children could be like. The [extracted] keywords ``earlier education'' really helped me to find the painpoint.''
\end{quote}
This feature helped P12 envision children’s daily life and translate it into actionable design opportunities. Similarly, P7 highlighted that personas could help designers address needs that might otherwise be missed.

Finally, participants cautioned that directly eliciting user preferences, as in the baseline system, could introduce risks. P7 emphasized that some user-generated suggestions may be impractical or even harmful:
\begin{quote}
P7: ``I've seen design decisions where I'm like, are you trying to kill our clients? [...] We asked the kids [on baseline system] what they wanted, they asked you for something dangerous [swivel chair], right? Like, that's… that's why we don't do that, right? Like, if you ask a kid what they want to eat for breakfast, they'll tell you cake, and then get diabetes and die, right?''
\end{quote}
This tension highlights how \tool’s persona-driven approach can reduce barriers to ideation while mitigating the risks of relying solely on direct user input.


\subsubsection{Collaborative ideation with personas}
Several participants (P1, P2, P4, P8, P10) emphasized that \tool could support more collaborative forms of ideation by helping to align designers, clients, and target users around a shared understanding of the persona. In practice, participants often experienced tension between how clients imagined their customers and the realities of user needs. P10, for instance, described how clients frequently start projects with a preconceived aesthetic and then retroactively fit an `ideal customer' onto that vision. In such cases, \tool was seen as a way to ground discussions in persona-driven evidence and refine ideation more systematically:  

\begin{quote}
P10: ``So yeah, anyway, typically here, I would have a client who's already got a decent idea of what they want. They've already got an idea of who the customer is. They generally have an idea of an aesthetic, and then they've retrofitted their ideal customer onto that. I see. And usually, usually you have to sort of talk them out of that. So [...] I would be using this [\tool] to refine the ideation process.''
\end{quote}

While P10 emphasized using personas to recalibrate client expectations, P1 saw opportunities for directly involving end users in the ideation process. They suggested that the tool could be configured with persona attributes and then shared with representative users to gather early signals of preference through visual selections. This process, P1 argued, could extend collaboration beyond the design team, building continuity between user feedback, persona representations, and the evolving prototype:  

\begin{quote}
P1: ``The tool [\tool], I will just bring it to the users. For example, I will do the filter preset for them. [...] And I let them pick the stuff they're looking for. If we have, like, target groups, I will bring the computer to them, and then I'll just show, oh. What kind of stuff you like? Like, I have these images, like, can you pick one, or can you vote which one you're looking for? It doesn't necessarily need to be, like, a lot, but at least I can have a base ideas of what, these, our customers are looking for.''
\end{quote}

P1 further highlighted how such collaborative engagement could reduce costly misalignments later in the design process. By bringing clients and users into the ideation stage through persona-linked products, designers could ensure that outcomes felt co-created rather than imposed:  

\begin{quote}
P1: ``If we have an AI tools that, can bridge, like, our customers, the designer and clients. The [client] companies, they will be, huge helps. Yeah, [it can reduce] a lot [of time], because, yeah, because we only cooperate from the very beginning. And everything is here, everything is visualized. So they… when we come out with the final design, they don't feel surprised. ``Oh, why you come up with stuff like this? It's not what we discussed'', right? It's like, something we worked on all along from the beginning.''
\end{quote}

Together, these perspectives illustrate how \tool could enable new forms of collaborations grounded in the needs of target populations: by helping designers negotiate client assumptions and by enabling more participatory engagement without incurring upfront cost of identifying and conducting research with target users too early in the process. In the following section, we discuss potential opportunities for co-designing with virtual personas before or after engaging with representative real world populations.



%


\section{Discussion}
In this section, we interpret our findings in light of both opportunities and risks associated with AI-generated personas in product design. We then reflect on the limitations of our study and how they shape the scope of our conclusions. Finally, we outline directions for future research, focusing on how persona-driven tools might be extended to more robustly consider and capture preferences of target consumers.

\subsection{Opportunities and Risks of AI-Persona Use in Product Ideation}
Our findings demonstrate the potential for integrating AI-generated personas as early-stage ideation tools within design workflows. These virtual personas expand designers' access to diverse customer profiles and provide contextualized visual stimuli that can spark new design directions. Yet, as prior work cautions, this promise is accompanied by risks. Kaate et al. showed that participants often accepted fabricated persona responses, reflecting a tendency to overtrust AI-generated content~\cite{kaate2025personas}. This overtrust and overreliance underscores a growing concern in HCI and machine learning communities regarding virtual agents~\cite{amin2025generative,shin2024understanding,sun2024building}. While researchers have proposed grounding personas in verifiable, human-derived data (\eg census distributions~\cite{castricato2024persona,li2025llm}, surveys~\cite{jung2025personacraft,kaate2025you}, consumer behavior~\cite{berke2024insights}, product/service reviews~\cite{clements2023innovative} or opinion polls~\cite{castricato2024persona}), to mitigate some of these risks, such approaches are costly, quickly outdated, and difficult to scale.

Our study offers complementary opportunities to address these risks by increasing the transparency of system reasoning. Participants highlighted that \tool's multimodal workflow, progressing from persona descriptions (\textbf{text}) \textrightarrow product exemplars (\textbf{images}) \textrightarrow extracted features (\textbf{text}) \textrightarrow outcome visualizations (\textbf{images}), created a more transparent and interpretable process for validation. \revised{This structured cross-modal pipeline allowed designers to trace how system outputs emerged, specifically addressing the inherent ambiguity of natural language. While subjective adjectives like \textit{rugged}, \textit{durable}, or \textit{adjustable} often evoke divergent mental models in isolation,~\tool anchored these terms in the persona's specific context. By extracting these textual descriptors directly from visual product exemplars that were rooted in the persona's profile, the system established a visible lineage between abstract persona traits and concrete design aesthetics. This enabled designers to verify that the generated visual cues were not arbitrary model hallucinations, but derived from the persona's established lifestyle and preferences. This visibility fundamentally altered user behavior: rather than relying on opaque and time-consuming trial-and-error prompting, participants shifted toward iterative exploration of the persona and its derivatives, such as products and features.}
\revised{Together, these insights suggest that structured, interpretable workflows can mitigate risks of overtrust by making system reasoning visible and grounding outputs in traceable references. Our results indicate that the reliability of AI personas in professional workflows depends as much on the transparency and traceability of the interaction process with AI as it does on the underlying model performance. This finding challenges the assumption that increasing realism is the only path to utility, raising broader questions about the alignment between interaction modalities and design goals, which we discuss further in \autoref{dis:fidelity}.}

\subsection{\revised{Beyond Fidelity: Interaction and Task Alignment with AI Personas}}
\label{dis:fidelity}

\revised{While achieving precise representation in AI personas remains an ongoing challenge in the machine learning community, our development of~\tool raises a fundamental question: \textit{is a fully rich and hyper-realistic representation of AI personas truly required for early-stage product design ideation?}}

\revised{Fundamentally, product design utilizes personas to represent \textit{archetypes}--groups sharing common traits and needs--rather than a single specific individual~\cite{chang2008personas}. However, standard conversational interfaces (chat) inherently frame the interaction as role-playing~\cite{shanahan2023role}. This dyadic structure encourages designers to probe for specific demographic details, creating an illusion that they are conversing with a distinct and stable individual (\eg P4, P8). This creates friction with the underlying technology; as P8 observed, the non-deterministic nature of LLMs means the same persona may respond differently in subsequent sessions. When a chat interface frames the AI as a specific person, this inconsistency breaks the user's suspension of disbelief, undermining trust in the system's reliability. Therefore, the issue is not solely whether the AI can statistically represent a population, but how the interface mediates that representation. Even if an AI model were perfectly representative of a demographic, the chat modality forces a conversational dynamic that may be ill-suited for extracting high-level archetypal insights.}

\revised{This perspective suggests that we do not see current representational limitations of AI personas as a fatal blocker for system design. While the pursuit of higher-fidelity models remains important for realistic simulation, we argue that the utility of AI personas today depends on aligning the interaction modality with the design phase. Conversational interfaces may remain the gold standard for post-hoc evaluation, where a designer wishes to user test a developed product against a simulated reaction (\eg Proxona~\cite{choi2025proxona}, UXagent~\cite{lu2025uxagent}). However, in the \textit{early conceptualization phase}, where the goal is to explore broad traits rather than simulate specific reactions, chat-based interfaces may inherently misdirect the designer's focus. Structured multimodal interfaces that enforce specific interaction constraints, could offer a way to bypass representational hurdles by abstracting the persona back into a useful design tool, rather than presenting it as an role-playing partner.}

\subsection{Limitations and Future Work}
Our study has several limitations that contextualize the scope of our findings. First, our user study involved a small sample size and was limited to professional designers focused on physical products, who may not represent the diversity of design domains or practices more broadly. Second, the task procedure was capped at 30 minutes, which may have unintentionally shaped participant behaviors that diverge from naturalistic design practices. Third, we compared \tool to a free-form LLM baseline, which may not reflect all the ways in which designers use personas in their work. Additionally, the baseline system did not include comparable features such as the ability to search for images online (\eg Pinterest), which may have disadvantaged it as a point of comparison. Lastly, because the study focused on short-term sessions, the findings may not generalize to long-term or collaborative design practices where personas are used iteratively across project stages and organizational contexts. Addressing these limitations will require future studies situated in ecologically valid settings and spanning a broader range of domains. Despite these constraints, our findings point to meaningful opportunities for supporting designers in using personas as active tools that shape design reasoning and outcomes.

Central to our system is the use of product references: visual artifacts that help designers reason about how a persona might perceive or interpret a given design. Unlike abstract attributes (\eg \textit{minimalist aesthetics}), product references offer grounded proxies for taste and expectation. However, reliance on product references alone poses risks in the absence of up-to-date consumer trends data; these references may introduce misplaced confidence in recommendations that do not reflect current market preferences. In this work, we leaned on LLMs' ability to continuously ingest large-scale data streams, including consumer trends, to mitigate some of these risks. However, without continuous validation against evolving consumer trends, references may instill misplaced confidence in outdated or unrepresentative exemplars, especially in fast-moving markets.

While out of the scope of this paper, generating and validating such references remains a challenging problem, particularly in physical product domains where obtaining reliable, domain-general ground truth is either too costly or infeasible. Recent work in information retrieval and preference modeling offer promising directions, including using consumer behavior data to construct richer personas (\eg \textit{busy parents}), thereby improving the relevance and scalability of product recommendations in e-commerce applications~\cite{wang2025opera,wang2024towards,shi2025you} Complementary approaches on pluralistic alignment and few shot preference optimization techniques further support this goal by integrating diverse customer signals~\cite{singh2025fspo,zhang2024guided,ryan2025synthesizeme,hosseini2025retrieve}. Together, these advances promise a path toward generating product references that continue to evolve alongside shifting markets and consumer preferences, and effectively serve as creative stimuli and points of validation in persona-driven design workflows. 

\section{Conclusion}
This paper introduced \tool{}, a multimodal LLM system that transforms personas from static artifacts into dynamic, actionable tools for design. Our study with professional designers shows promising potential for \tool{} to improve transparency, satisfaction, and engagement compared to a persona $+$ chat-based interaction baseline, while revealing tensions between structured support and open-ended exploration. Future work should extend such systems into collaborative and longitudinal workflows to better align design imagination, stakeholder expectations, and evolving user needs.


\bibliographystyle{ACM-Reference-Format}
\bibliography{references}

\appendix
\section{Implementation Details}
\subsection{System Prompts}
\label{appx:prompt}

\begin{itemize}
    \item Product inference (\twemoji{bulb})
\end{itemize}

\begin{lstlisting}[style=prompt]
Given <persona> below, what kind of physical products would be on the wishlist for their birthday or Christmas presents?

<persona> ${persona} </persona>

### general_guideline ###

NEVER use meta-phrases (e.g., `let me help you', `I can see that').
NEVER use phrases like `I am an AI language model'.
NEVER use phrases like `I am not able to'.
NEVER include any preambles or introductory remarks. Respond immediately and directly.


### task_instruction ###

List exactly 9 distinct physical products, each from a different **top-level** physical product category. For each item, specify a real, existing **brand** and **model name**.
Include a diverse range of categories of physical products like gadgets, appliances, furnitures, tools, considering various relatable situations and motivations such as home, work/study, mobility, leisure, health/wellness, hobbies/creative, caregiving, social, and outdoor. Use at most one item per top-level category of physical products. **NEVER** include non-physical products such as membership, map, class, food, book, music, movie, content, insurance, and subscription services.
Then, explain in detail why the <persona> would like to purchase or receive each item for their special gift, carefully considering their unique and specific persona characteristics.
When reasoning, comprehensively consider both <objective_attributes> and <subjective_attributes> to select products and brands that plausibly align with what <persona> wishes to have or receive.
Structure your response exactly as described in <response_structure> below:


### persona_attributes ###

<objective_attributes>
age, sex, race, ancestry, education, employment_status, industry_category, occupation_category, income, disability
</objective_attributes>

<subjective_attributes>
big_five_scores: {openness, conscientiousness, extraversion, agreeableness, neuroticism}, defining_quirks, mannerisms, personal_time, lifestyle
</subjective_attributes>


### response_structure ###

<response_structure>
products: ["product1_name", "product2_name", "product3_name", "product4_name", "product5_name", "product6_name", "product7_name", "product8_name", "product9_name"]

attributes: [
"[persona_attributes_considered_for_product1], [persona_attributes_considered_for_product2], [persona_attributes_considered_for_product3], [persona_attributes_considered_for_product4], [persona_attributes_considered_for_product5], [persona_attributes_considered_for_product6], [persona_attributes_considered_for_product7], [persona_attributes_considered_for_product8], [persona_attributes_considered_for_product9]"
]

reasons: [
"product1_reasons_with_persona_attributes", "product2_reasons_with_persona_attributes", "product3_reasons_with_persona_attributes", "product4_reasons_with_persona_attributes", "product5_reasons_with_persona_attributes", "product6_reasons_with_persona_attributes", "product7_reasons_with_persona_attributes", "product8_reasons_with_persona_attributes", "product9_reasons_with_persona_attributes"
]
</response_structure>
\end{lstlisting}

\begin{itemize}
\item Feature extraction (\twemoji{jigsaw})
\end{itemize}

\begin{lstlisting}[style=prompt]

### general_guideline ###

NEVER use meta-phrases (e.g., `let me help you', `I can see that').
NEVER use phrases like `I am an AI language model'.
NEVER use phrases like `I am not able to'.
NEVER include any preambles or introductory remarks. Respond immediately and directly.


### task_instruction ###

Analyze and elicit product attributes of the img_url with respect to the <structure> below. Carefully consider ###definition### when analyzing the image. Consider both conventional and unconventional attributes for each key.

### definition ###

aesthetics: the sensation that results from the perception of attractiveness (or unattractiveness) in products. (e.g., color, material, shape)
behavioral: what a product is seen to say about its function, mode-of-use and qualities. (e.g., physical_action_possibilities, function, physical_interaction)
contextual: the perception of what a product says about its owner or user: the personal and social significance attached to the design. (e.g., symbolic, environment, situation)

### examples ###
For a helmet:
- behavioral:
    - conventional: ["wearing", "protecting"]
    - unconventional: ["hammering", "breaking", "throwing"]
- contextual:
    - conventional: ["riding a bike", "safety"]
    - unconventional: ["kitchen", "party"]

NEVER make any sentences with attributes.

### output_structure ###
<structure>
aesthetics: [color: color_feature1, color: color_feature2, color: color_feature3, material: material_feature1, material: material_feature2, material: material_feature3, shape: shape_feature1, shape: shape_feature2, shape: shape_feature3]
behavioral: [physical_action_possibilities: physical_action_possibilities_feature1, physical_action_possibilities: physical_action_possibilities_feature2, physical_action_possibilities: physical_action_possibilities_feature3, function: function_feature1, function: function_feature2, function: function_feature3, physical_interaction: physical_interaction_feature1, physical_interaction: physical_interaction_feature2, physical_interaction: physical_interaction_feature3]
contextual: [symbolic: symbolic_feature1, symbolic: symbolic_feature2, symbolic: symbolic_feature3, environment: environment_feature1, environment: environment_feature2, environment: environment_feature3, situation: situation_feature1, situation: situation_feature2, situation: situation_feature3]
</structure>

***IMPORTANT*** You MUST provide EXACTLY 9 unique and distinct attributes for each dimension, that is, 9 aesthetics, 9 behavioral, and 9 contextual. Each attribute should be a phrase, more than just a single word.

\end{lstlisting}

\subsection{Attribute Hierarchy}
\label{appx:attribute}

\begin{itemize}
  \item Persona Filters
    \begin{itemize}
      \item Basic Info
        \begin{itemize}
          \item Age
            \begin{itemize}
              \item 18--25
              \item 26--35
              \item 36--45
              \item 46--55
              \item 56--65
              \item 65+
            \end{itemize}
          \item Sex
            \begin{itemize}
              \item Male / Female
            \end{itemize}
          \item Ancestry
            \begin{itemize}
              \item European/White
              \item African/African American
              \item Asian/Pacific Islander
              \item Hispanic/Latino
              \item Native American/Indigenous
              \item Middle Eastern/North African
              \item Mixed/Multiracial
              \item Other/Unspecified
            \end{itemize}
          \item Marital Status
          \item Household Type
            \begin{itemize}
              \item Living alone
              \item Married couple without children
              \item Married couple with children
              \item Cohabiting couple without children
              \item Cohabiting couple with children
              \item Single parent with children
              \item Living with parents/family
              \item Living with roommates/non-relatives
              \item Multi-generational household
            \end{itemize}
          \item Family Presence and Age
          \item Disability
        \end{itemize}
      \item Occupation
        \begin{itemize}
          \item Education
          \item Employment Status
            \begin{itemize}
              \item Employed
              \item Military/Armed Forces
              \item Unemployed
              \item Not in Labor Force
              \item Student
              \item Too Young to Work
              \item Other
            \end{itemize}
          \item Occupation Category
          \item Industry Category
            \begin{itemize}
              \item Administration \& Government
              \item Agriculture \& Forestry
              \item Arts \& Entertainment
              \item Construction
              \item Education
              \item Finance \& Insurance
              \item Healthcare \& Medical
              \item Technology \& Data
              \item Legal Services
              \item Manufacturing
              \item Military
              \item Professional Services
              \item Retail \& Sales
              \item Transportation \& Logistics
              \item Utilities
              \item Retired
              \item Not Applicable
            \end{itemize}
          \item Income
            \begin{itemize}
              \item No income
              \item \$0 -- \$25,000
              \item \$25,001 -- \$50,000
              \item \$50,001 -- \$75,000
              \item \$75,001 -- \$100,000
              \item \$100,001 -- \$150,000
              \item \$150,001+
            \end{itemize}
        \end{itemize}
      \item Personality (Big-Five)
        \begin{itemize}
          \item Openness (Low / Medium / High)
          \item Conscientiousness (Low / Medium / High)
          \item Extraversion (Low / Medium / High)
          \item Agreeableness (Low / Medium / High)
          \item Neuroticism (Low / Medium / High)
        \end{itemize}
      \item Behavioral Traits
        \begin{itemize}
          \item Communication Style
            \begin{itemize}
              \item Direct
              \item Diplomatic
              \item Casual
              \item Formal
              \item Expressive
              \item Reserved
            \end{itemize}
          \item Activity Style
            \begin{itemize}
              \item Active
              \item Relaxed
              \item Social
              \item Solitary
              \item Creative
              \item Routine
            \end{itemize}
          \item Social Style
            \begin{itemize}
              \item Community-oriented
              \item Family-focused
              \item Independent
              \item Social
              \item Private
              \item Professional
            \end{itemize}
          \item Defining Quirks
          \item Mannerisms
          \item Personal Time
          \item Lifestyle
        \end{itemize}
      \item Additional Info
        \begin{itemize}
          \item Household Language
          \item Citizenship
        \end{itemize}
    \end{itemize}
\end{itemize}

\section{Interaction Logs}
\label{appx:logs}
\begin{table*}[h!]
\centering
\caption{User logs showing the number of image generation, clicks of exploration buttons (\twemoji{bulb}, \twemoji{heavy_plus_sign}, \twemoji{jigsaw}), switching tabs (canvas or chat vs. production), and chat queries that users sent to LLM in the baseline.}
\begin{tabular}{l r !{\vrule width 0.5pt} r r r c !{\vrule width 0.5pt}  r c r r r r c c}
\toprule
    & \multicolumn{7}{c}{Personagram} & \multicolumn{4}{c}{Baseline}
    \\ \cmidrule(lr){2-8} \cmidrule(lr){9-12} 
    
    PID & Generation & \twemoji{bulb} & \twemoji{heavy_plus_sign} & \twemoji{jigsaw} &  & Switch & Duration & Generation & Queries & Switch & Duration\\
    \midrule
     P1 & 5 & 2 & 1 & 5 & & 7 & 18.2 & 16 & 5 & 21 &29.6 \\ 
     P2 & 12 & 4 & 3 & 3 & & 3 &20.0 & 7 & 4 & 1 &13.1 \\ 
     P3 & 10 & 11 & 2 & 5 & & 9 &30.1 & 24 & 3  & 4 &22.4 \\ 
     P4 & 14 & 9 & 2 & 9 & & 9 &17.3 & 17 & 14 & 5 &23.7 \\ 
     P5 & 27 & 4 & 2 & 7 & & 11 &26.3 & 17 & 13 & 21 &27.1 \\ 
     P6 & 6 & 9 & 5 & 6 & & 13 & 14.2 & 9 & 2 & 3 & 13.1  \\ 
     P7 & 9 & 13 & 4 & 8 & & 7 & 12.4 & 19 & 7 &  12 & 25.6 \\ 
     P8 & 3 & 3 & 0 & 9 & & 15 & 14.8 & 8 & 12 & 11 & 19.8 \\ 
     P9 & 6 & 4 & 0 & 7 & & 15 & 11.8 & 8 & 7 &  19 & 17.8 \\ 
    P10 & 8 & 4 & 2 & 4 & & 11 & 17.4 & 5 & 4 & 5 & 12.6 \\ 
    P11 & 28 & 8 & 5 & 16 & & 21 & 29.4 & 14 & 13 & 15 & 32.6  \\ 
    P12 & 13 & 5 & 2 & 4 & & 9 & 22.5 & 9 & 8 & 9 & 23.9 \\
    \midrule
    Total & 141 & 76 & 28 & 83 & (187) & 130 & 19.6 (6.2) & 153 & 92 & 126 & 21.8 (6.6) \\
\bottomrule
\end{tabular}
\label{tab:userlogs}
\end{table*}

\section{Analysis}
\label{appx:analysis}
\begin{table*}[ht]
\centering
\caption{Frequency counts of LLM prompt queries organized by themes, with representative examples. *Counts are not mutually exclusive.}
\begin{tabular}{llp{5.5cm}r}
\hline
\textbf{Category} & \textbf{Subcategory} & \textbf{Example Query} & \textbf{Count} \\
\hline
\multirow{2}{*}{A. Persona Exploration} 
  & A1. Persona attributes & \textit{``How tall is Diego?''} & 6  \\
  & A2. Persona needs/preferences & \textit{``What colors would they collectively agree on?''} & 32 \\
\hline
\multirow{4}{*}{B. Concept Exploration} 
  & B1. Product ideation & \textit{``Design a chair as a compact playground.''} & 20 \\
  & B2. Aesthetic & \textit{``Make the design more modern and tasteful.''} & 17 \\
  & B3. Functional & \textit{``Focus on weak eyesight, hearing, safety first.''} & 47 \\
  & B4. Symbolic & \textit{``What chairs would help girls feel welcome?''} & 7 \\
\hline
\multirow{2}{*}{C. Prototyping} 
  & C1. Prompt generation & \textit{``Give me a text-to-image helmet prompt.''} & 24 \\
  & C2. Prompt refinement & \textit{``Rewrite into four short sentences.''} & 20 \\
\hline
\end{tabular}
\end{table*}









\end{document}